%% file: main.tex
\documentclass[twocolumn]{aastex63}
\usepackage{amsmath, mathtools}
\usepackage{courier}
\usepackage{url}

\shorttitle{}
\shortauthors{Claytor et al.}

\newcommand{\sol}{\odot}
\newcommand{\Teff}{T_{\mathrm{eff}}}
\newcommand{\logg}{\log{g}}
\input{numbers.tex}
\input{numbers_jvs.tex}

\begin{document}

\title{Chemical Evolution in the Milky Way: Rotation-based ages for APOGEE-\textit{Kepler} cool dwarf stars}

\author[0000-0002-9879-3904]{Zachary R. Claytor}
\affiliation{Institute for Astronomy, University of Hawai'i, 2680 Woodlawn Drive, Honolulu, HI 96822, USA}

\author[0000-0002-4284-8638]{Jennifer L. van Saders}
\affiliation{Institute for Astronomy, University of Hawai'i, 2680 Woodlawn Drive, Honolulu, HI 96822, USA}

\author[0000-0001-7195-6542]{\^{A}ngela R. G. Santos}
\affiliation{Space Science Institute, 4750 Walnut Street, Suite 205, Boulder, CO 80301, USA}

\author[0000-0002-8854-3776]{Rafael A. Garc\'{i}a}
\affiliation{IRFU, CEA, Universit\'e Paris-Saclay, F-91191 Gif-sur-Yvette, France}
\affiliation{AIM, CEA, CNRS, Universit\'e Paris-Saclay, Universit\'e Paris Diderot, Sorbonne Paris Cit\'e, F-91191 Gif-sur-Yvette, France}

\author[0000-0002-0129-0316]{Savita Mathur}
\affiliation{Departamento de Astrof\'isica, Universidad de La Laguna, E-38206 Tenerife, Spain}
\affiliation{Instituto de Astrof\'isica de Canarias, C/ V\'ia L\'acteer s/n, La Laguna, E-38205 Tenerife, Spain}

\author[0000-0002-4818-7885]{Jamie Tayar}
\altaffiliation{Hubble Fellow}
\affiliation{Institute for Astronomy, University of Hawai'i, 2680 Woodlawn Drive, Honolulu, HI 96822, USA}

\author{Marc H. Pinsonneault}
\affiliation{Department of Astronomy, The Ohio State University, 140 West 18th Avenue, Columbus, OH 43210, USA}

\author{Matthew Shetrone}
\affiliation{University of Texas at Austin, McDonald Observatory, 32 Fowlkes Rd, McDonald Observatory, TX 79734-3005, USA}

\date{\today}

\begin{abstract}
    We use models of stellar angular momentum evolution to determine ages for $\sim500$ stars in the APOGEE-\textit{Kepler} Cool Dwarfs sample. We focus on lower main-sequence stars, where other age-dating tools become ineffective. Our age distributions are compared to those derived from asteroseismic and giant samples and solar analogs. We are able to recover gyrochronological ages for old, lower-main-sequence stars, a remarkable improvement over prior work in hotter stars. Under our model assumptions, our ages have a median relative uncertainty of 14\%, comparable to the age precision inferred for more massive stars using traditional methods. We investigate trends of galactic $\alpha$-enhancement with age, finding evidence of a detection threshold between the age of the oldest $\alpha$-poor stars and that of the bulk $\alpha$-rich population. We argue that gyrochronology is an effective tool reaching ages of 10--12 Gyr in K- and early M-dwarfs. Finally, we present the first effort to quantify the impact of detailed abundance patterns on rotational evolution. We estimate a $\sim15\%$ bias in age for cool, $\alpha$-enhanced (+ 0.4 dex) stars when standard solar-abundance-pattern rotational models are used for age inference, rather than models that appropriately account for $\alpha$-enrichment.
\end{abstract}

\section{Introduction}
Our understanding of Milky Way formation and evolution is informed by kinematics and chemistry of different stellar populations. Studies of galactic archaeology \citep[e.g.,][]{Edvardsson93, Feltzing98, Bensby03, Haywood13, Bensby14, Feltzing17} have constructed the current schematic of our galaxy's disk: an old, ``thick" disk poor in metals but enhanced in $\alpha$-process elements; and a young, metal-rich, $\alpha$-poor ``thin" disk. Kinematically, the thick disk is characterized by high velocity dispersion perpendicular to the galactic plane, while thin-disk stars remain near the plane. Trends of composition with age suggest these populations underwent separate phases of chemical enrichment, thought to be driven by different kinds of supernovae \citep{Maoz11}. These trends have even led some to suggest certain abundances as proxies for age \citep[e.g.,][]{Bensby14, Martig16, TucciMaia16, Feltzing17}. Other studies have argued against the thin- and thick-disk model of the Milky Way: as our ability to estimate precise stellar ages, composition, and kinematics has improved, the historical two-population hypothesis has evolved into a continuum of galactochemical structure and evolution \citep[e.g.,][]{Bovy12a, Bovy12b, Buder19}.

The study of chemical evolution in stellar populations is impossible without precise ages. To date, most investigations have used isochrones to estimate ages \citep{Nordstrom04, Haywood13, Buder19}. While easy to implement, isochrone ages are most useful for cluster stars, which provide an ensemble of stars at a single age, or for field stars that have aged to about one-third of their main-sequence lifetimes \citep{Pont04, Soderblom10} and thus move substantially on the Hertzsprung-Russell (H-R) diagram. In this regime, isochrone ages are, at best, precise to 15\% \citep[e.g.,][]{Nordstrom04}. However, this excludes the majority of stars, which are not massive enough to have evolved sufficiently within the age of the galaxy. For the lowest-mass stars, which are practically stationary in temperature--luminosity space, isochrones provide no age constraints whatsoever \citep{Pont04, Epstein14}.

Asteroseismology can also provide valuable constraints on global stellar properties. Virtually all cool stars excite solar-like oscillations, with timescales ranging from minutes to months depending on the mean density of the star. These oscillation frequencies can be measured successfully for large samples of stars using time-domain space data \citep[e.g.,][]{Stello13, Hon19, Schofield19}. The frequency of maximum power is related to the surface gravity \citep{Kjeldsen95}, and the frequency spacing between modes is related to the mean density via asymptotic pulsation theory. This information can be combined to infer mass and radius for large samples of stars. For red giants, knowledge of mass and composition alone gives valuable age information \citep[e.g.,][]{SilvaAguirre18}; mass information also allows more precise ages for solar analogs than one can obtain from H-R diagram position alone. By analyzing the frequency pattern in detail, one can also obtain more direct age constraints tied to helium production in the core, giving fractional age uncertainties as low as 5\% in the best cases \citep{Creevey17}. Again, however, for the lower main sequence these asteroseismic signals become ineffective chronometers because age evolution at fixed mass produces small changes in either central helium abundance or radius. Furthermore, asteroseismic signals become difficult to detect in stars below solar mass. Studies of galactic dwarfs require a more uniformly accessible chronometer.

With the advent of large time-domain photometric surveys, increasingly large samples of stellar rotation periods have become available \citep[e.g.,][who detected periods in over 80\% of late-K and early M-stars in the \textit{Kepler} field]{McQuillan14}. These period estimates are natural byproducts of transit and transient surveys and enable the inference of ages using gyrochronology \citep{Barnes03, Barnes07}, based on the observation that main-sequence stars spin down as they age \citep[e.g.,][]{Skumanich72}. \citet{Barnes07} argued that with proper calibration, gyrochronology could be used to infer stellar ages to within 15\%, requiring knowledge only of the color (a proxy for mass) and rotation period. This promise has driven several investigations of rotation--age relationships, typically focused on cool, main-sequence stars \citep[e.g.,][]{Barnes07, Mamajek08, Barnes10, Epstein14, Angus15, Gallet15, Meibom15, vanSaders16}.

Using the spin-down law of \citet{Krishnamurthi97}, \citet{Epstein14} explored the strengths and weaknesses of rotation-based methods, concluding that gyrochronology may be a more precise clock than both asteroseismology and isochronal techniques for cool main-sequence stars. It has potential for Sun-like stars as well: \citet{Meibom15} obtained rotation periods for stars in NGC 6819, a 2.5 Gyr-old cluster, and demonstrated that gyrochronology could be used to obtain $\sim 10\%$ ages for solar-temperature stars.

Gyrochronology comes with challenges as well. Purely empirical methods are restricted to the domain in which they are directly calibrated, which can severely limit their utility. Rotational properties are strong functions of mass and evolutionary state, making extrapolation to regimes with poorly-characterized period-mass-age relationships risky. Theoretical models can, in principle, allow more robust predictions for stars not directly constrained by data. However, the underlying physical model can be complex. We outline the strengths and weaknesses of these different approaches below.

Empirical approaches can be powerful and flexible, using a methodology similar to that employed by color-temperature relationships: one takes calibrators of known age and correlates observables, such as color and rotation period, with age. Calibration sources tend to be open clusters \citep[modern \textit{Kepler}-era calibrators:][]{Meibom11, Barnes16, Hartman09, Agueros18, Rebull16, Rebull17, Douglas16}, asteroseismic stars \citep{Garcia14b, Angus15, vanSaders16}, or binaries for which individual periods are known \citep{Mamajek08, Chaname12}. These come with important systematics. For example, old clusters are generally more distant than young ones. As a result, calibration samples for old, lower-main-sequence stars are sparse due to the challenges of observing those distant clusters. Amplitudes of variation are smaller for hot stars and slow rotators, so rotational detection is biased against older and more massive stars as well. There are also few calibrators of non-solar metallicity. Consequently, the composition dependence of stellar spin-down remains poorly constrained, and the behavior at older ages is subject to severe observational selection effects. More generally, structural evolution (through angular momentum conservation) can induce large departures from simple power-law behavior \citep[see][for a discussion on subgiant rotation]{vanSaders13}.

Theoretical gyrochronology models require choices for the torque associated with magnetized winds, angular momentum transport, and initial rotation conditions. While braking laws \citep[e.g.,][]{Kawaler88, Krishnamurthi97, Matt08, Matt12, vanSaders13} are physically and empirically motivated, they require some assumptions regarding magnetic field strength, geometry, and stellar mass loss. Many spin-down relations treat stars as solid bodies. While this assumption is reasonable for old Sun-like stars, both theoretical \citep{Gallet15, Denissenkov10} and observational \citep{Curtis19, Agueros18} investigations suggest that the neglect of internal angular momentum transport is a poor assumption in very young objects. Furthermore, a single set of initial conditions is often chosen to represent all stars, but the period distributions of young cluster members demonstrate both striking mass-dependent trends and a wide range of initial rotation rates at fixed mass \citep{Hartman09, Rebull16}. While the braking laws predict that these initial conditions matter less for older stars \citep{vanSaders13, Epstein14}, there are combinations of stellar masses and ages for which this initial scatter should not be ignored.

There is also evidence that the underlying model can break down for inactive stars. \citet{vanSaders16} found that stars more than halfway through their main-sequence lifetimes experience weakened braking \citep[see also][]{Angus15}. This poses extra constraints on the types of stars that can safely be considered in gyrochronological studies. 

Stars less massive than the Sun have longer main-sequence lifetimes and thus take longer to reach the main-sequence halfway point. The coolest K- and M-dwarfs cannot have reached this point within the age of the galactic disk, so weakened braking should not pose a problem for the lowest-mass stars. In this work we focus on these low-mass stars, for which we believe gyrochronology has the greatest potential. We adopt a forward modeling approach, taking advantage of the ability of theoretical models to account for the effects of composition on angular momentum evolution predictions. We obtain rotation-based ages for a sample of cool dwarf stars observed by both \textit{Kepler} and the Apache Point Observatory Galactic Evolution Experiment \citep[APOGEE,][]{Zasowski13, Majewski17}. Using our derived ages, we explore the evolution of $\alpha$-element abundances in the Milky Way and attempt to recover chemical evolution trends seen in the literature. We investigate the biases inherent to the detection of rotation periods and consequently to age inference. Finally, we examine the sensitivity of our age estimates to assumptions of metallicity, $\alpha$-abundance, and initial rotation conditions, demonstrating the viability of gyrochronology as a tool for astrophysical problems.

\section{Methods}
\label{sec:Methods}
The use of model-based gyrochronology requires several assumptions. First are choices of the actual input physics of stellar evolution, including chemical evolution and the treatment of convection. Invoking rotation forces a choice of internal angular momentum transport, braking law, and initial conditions. While our model-driven approach employs physically motivated choices for all these, each choice still requires empirical calibration based either on the Sun alone or on a small sample of well-studied stars. We explore these choices in detail below. Because of the assumptions built into our method, readers should note that 1) our results are unavoidably model-dependent, and that 2) making different model choices will affect the absolute ages, but the ranked order should be preserved. Understanding the order of galactic formation requires accurate relative ages, but accuracy in absolute age is not as necessary.

\subsection{Stellar Spin-Down}
Current theory of stellar spin-down or braking uses charged stellar winds in a magnetic field to carry angular momentum away from the star. The most basic braking laws have the form $dJ/dt \propto \omega^3$ \citep{Kawaler88} for stars of a given mass, where $J$ is the angular momentum, and $\omega$ is the angular rotation velocity. In other words, faster-rotating stars spin down more quickly, while more slowly-rotating stars spin down more slowly. This means that stars with similar masses born at nearly the same time will asymptotically approach the same spin rate, regardless of their initial angular speeds. Thus we can map the rotation period of a star at late time robustly to its age, provided the mapping is properly calibrated. This relationship is valid for stars with $M \lesssim 1.3 M_\sol$ or $\Teff \lesssim 6200$ K. Above these regimes the surface convection zone vanishes, affecting the magnetic field generation and thereby the spin-down process \citep[this is known as the Kraft break, after][]{Kraft67}.

Describing the period distributions observed in clusters requires more complicated braking prescriptions than the Kawaler law. The discovery of rapidly-rotating stars in the Pleiades cluster \citep{Soderblom83, Stauffer84} suggested that angular momentum loss must be weaker in some regimes. Observations of x-ray emission, a proxy for magnetic activity, also suggested that stars' magnetic fields saturate above some critical rotation speed $\omega_\mathrm{crit}$ \citep[e.g.,][]{Vilhu83, Wright11}. This saturation would allow for weaker magnetic braking for the fastest-rotating stars. Using magnetic saturation to modify the Kawaler law, \citet{Krishnamurthi97} and \citet{Sills00} were able to reproduce the period distributions in open clusters. This modified braking law has the form

\begin{equation} \nonumber
	\frac{dJ}{dt} =
	\begin{dcases}
		-K \omega_\mathrm{crit}^2 \omega \left( \frac{R}{R_\sol} \right)^\frac{1}{2} \left( \frac{M}{M_\sol} \right)^{-\frac{1}{2}}, ~\omega > \omega_\mathrm{crit} \\
	    -K \omega^3 \left( \frac{R}{R_\sol} \right)^\frac{1}{2} \left( \frac{M}{M_\sol} \right)^{-\frac{1}{2}}, ~\omega \leq \omega_\mathrm{crit}
	\end{dcases}
\end{equation}
where
\begin{equation} \nonumber
    \omega_\mathrm{crit} = \omega_{\mathrm{crit},\sol} \frac{\tau_\mathrm{cz}}{\tau_{\mathrm{cz},\sol}},
\end{equation}
$\tau_\mathrm{cz}$ is the convective overturn timescale, and $K$ is a constant calibrated so that the law reproduces the Sun's rotation at the solar age. $R$ and $M$ are the stellar radius and mass, respectively, and the subscript $\odot$ denotes solar values.

This form of the braking law ties the magnetic braking to the properties of the convection zone \citep[e.g.,][]{Wright11, Cranmer11}, introducing an additional mass dependence. However, to second order, metallicity also affects the properties of the convection zone: metallicity increases opacity, which increases the depth of the surface convection zone and convective overturn timescale. Furthermore, metal content impacts the main-sequence lifetimes of stars at fixed mass, meaning that stars of different metallicities have different amounts of time over which to lose angular momentum on the main sequence. Two stars with the same mass and age, but different compositions, will therefore have different rotation periods.

Empirical braking laws generally neglect metallicity because there are too few calibrators to span that dimension sufficiently. Most open clusters used to calibrate spin-down relations have solar metallicity, so using these relations to predict angular momentum loss for stars with non-solar abundances requires extrapolation. However, with a full evolutionary model, one can account for the impact of metallicity on stellar structure and lifetime naturally and then apply a braking law to the model. This will result in a metallicity-dependent spin-down relation, and while there is still a lack of calibrators beyond solar metallicity, the extrapolation to other metallicities is physically motivated. 

Using the braking law in \citet{Matt08} and \citet{Matt12}, \citet{vanSaders13} produced such a model-based spin-down relation. They presented a torque of the form

\begin{equation}
    \label{eq:spindown}
	\frac{dJ}{dt} =
	\begin{dcases}
		f_K K_M \omega \left( \frac{\omega_\mathrm{crit}}{\omega_\sol} \right) ^2, ~\omega_\mathrm{crit} \leq \omega \frac{\tau_{\mathrm{cz}}}{\tau_{\mathrm{cz},\sol}} \\
		f_K K_M \omega \left( \frac{\omega \tau_{\mathrm{cz}}}{\omega_\sol \tau_{\mathrm{cz},\sol}} \right) ^2, ~\omega_\mathrm{crit} > \omega \frac{\tau_{\mathrm{cz}}}{\tau_{\mathrm{cz},\sol}},
	\end{dcases}
\end{equation} 

with

\begin{equation} \nonumber
    \frac{K_M}{K_{M,\sol}} = R^{3.1} M^{-0.22} L^{0.56} P_{\mathrm{phot}}^{0.44}.
\end{equation}
Here $f_K$ is the solar calibration constant, and $R$, $M$, $L$, and $P_\mathrm{phot}$ are the radius, mass, luminosity, and photospheric pressure, respectively, in solar units. To summarize, \citet{vanSaders13} took the \citet{Matt08} formulation, which uses the magnetic field strength and mass loss rate, and parameterized them in terms of photospheric pressure, rotation velocity, and convective overturn timescale based on empirical relations \citep{Hartman09, Wood05, Pizzolato03}. Notice that, for a star of known mass, composition, and age, the other parameters in $K_M$ can be determined from a stellar model, allowing for extrapolation of the spin-down law to stars with non-solar composition. While this partially accounts for metallicity in the braking law, the exponents of the braking law parameters are anchored in empirical calibrations which are largely based on stars with Sun-like composition. We do not treat the composition dependence of these exponents, nor any metallicity dependence of the initial conditions, both of which may be important. Instead, we focus only on the underlying stellar evolution models. The remaining values of $f_K$ and $\omega_\mathrm{crit}$, as well as initial conditions (i.e., initial disk period and disk-locking timescale) are typically calibrated using solar and cluster data \citep{vanSaders13}.

\begin{deluxetable}{l|lll}
    \caption{YREC model grid nodes}
    \label{tab:YREC}
    \tablehead{& \colhead{min} & \colhead{max} & \colhead{step}}
    \startdata
    mass ($M_\odot$)   & 0.3  & 2.0 & 0.01  \\
    {[}M/H{]} (dex)    & -1.0 & 0.5 & 0.5  \\
    {[}$\alpha$/M{]} (dex)  & 0.0  & 0.4 & 0.4  \\ \hline
    \enddata
\end{deluxetable}

The braking law in Eq.~(\ref{eq:spindown}) has been tested in several regimes against other spin-down models. For example, \citet{vanSaders13} showed that this law reproduces the rapid rotation of stars near the low-mass side of the Kraft break better than other Kawaler law variants. Similarly, \citet{Tayar18} demonstrated the strength of Eq.~(\ref{eq:spindown}) in reproducing the rotation periods of intermediate-mass core-helium-burning stars. For stars more like the Sun, most braking laws perform equally well since they are all anchored to the same set of Sun-like calibration stars. Because of its track record in reproducing observed periods of stars of various masses and evolutionary stages, we adopt the braking law in Eq.~(\ref{eq:spindown}) for our study.

For our analysis we make use of the grid of stellar rotation models created using the Yale Rotating stellar Evolution Code \citep[YREC,][]{Demarque08} from \citet{vanSaders13} and \citet{vanSaders16}. These models do not intrinsically include rotation. Instead, rotational evolution is determined by integrating Eq.~(\ref{eq:spindown}), while the models are used to provide the time and metallicity dependence of the stellar parameters appearing in the braking law. Thus, our rotational evolution does not include the effects of rotational mixing nor evolutionary feedback from starspots, both of which may be important for the most active and rapidly rotating stars. We used the same input physics as \citet{vanSaders16}, but recomputed the model grid using the \citet{Castelli04} atmosphere tables, which include non-solar abundances of $\alpha$-elements (O, Ne, Mg, Si, S, Ar, Ca, and Ti). 

Our model grid spans the range of mass, metallicity, and $\alpha$-abundance detailed in Table~\ref{tab:YREC}, while each evolution track has been run to either the helium flash or 30 Gyr, whichever occurs first. The models utilize the nuclear reaction rates of \citet{Adelberger11} and chemical evolution of the form $Y = Y_\mathrm{primordial} + \frac{\Delta Y}{\Delta Z} Z$, with $\frac{\Delta Y}{\Delta Z} = 1.4$. To model convection, YREC employs mixing-length theory \citep[MLT,][]{Vitense53, Cox68} with no convective overshoot. We use opacities from the Opacity Project \citep{Mendoza07}, supplemented with low-temperature opacities of \citet{Ferguson05}, for a \citet{Grevesse98} solar mixture of elements, and the OPAL equation of state \citep{Rogers02}. A solar calibration sets the values of the mixing length parameter $\alpha = 1.86$, solar hydrogen mass fraction $X_\odot = 0.7089$, and solar metallicity $Z_\odot = 0.0183$. The rotational evolution uses solid-body rotation and the angular momentum loss of Eq.~(\ref{eq:spindown}), with $f_K = 6.575$, $\omega_\mathrm{crit} = 3.394\times 10^{-5}\ \mathrm{s^{-1}}$, disk-locking timescale $\tau_\mathrm{disk} = 0.281$ Myr, and initial disk period $P_\mathrm{disk} = 8.134$ days \citep[the ``fast-launch" conditions of][]{vanSaders13}. We calculate the convective overturn timescale using the local MLT convective velocity at one pressure scale height above the convective boundary. We note that the effective temperatures for our sample (see Sect.~\ref{sec:Sample}) are generally in a regime where the weakened braking observed by \citet{vanSaders16} should not be important; most of our stars are not massive enough to have evolved to half of their main-sequence lifetimes within the age of the galaxy. Our tracks have been condensed in the time-domain to a more practical set of equivalent evolutionary phases (EEPs) according to the method of \citet{Dotter16}.

\subsection{Sample}
\label{sec:Sample}
We study the sample of \textit{Kepler} Cool Dwarfs from APOGEE-1 \citep{Majewski17} as described in the target selection of \citet{Zasowski13}. APOGEE-1, one of the programs in phase three of the Sloan Digital Sky Survey \citep[SDSS-III,][]{Eisenstein11}, spectroscopically surveyed the entire Milky Way with resolution $R \sim 22,500$ in the near-infrared. While APOGEE was designed primarily to constrain dynamical and evolution models of our galaxy, in collecting data across all galactic regions, the survey also targeted the \textit{Kepler} field. The APOGEE-1 dataset therefore contains stars for which we also have rotation periods from \textit{Kepler}. The stars in our sample were specifically targeted as an ancillary program to the core APOGEE survey. They were selected using \textit{Kepler} Input Catalog \citep[KIC,][]{Brown11} parameters to have $\Teff \leq 5500$ K, $\logg \geq 4.0$ dex, and $7 < H$-magnitude $ < 11$. With this intentionally simple selection function, the target sample contained \allstars{} stars. However, three plates were not observed as planned, so only \obsstars{} were observed. 

\begin{figure}
    \centering
    \includegraphics[width=\linewidth]{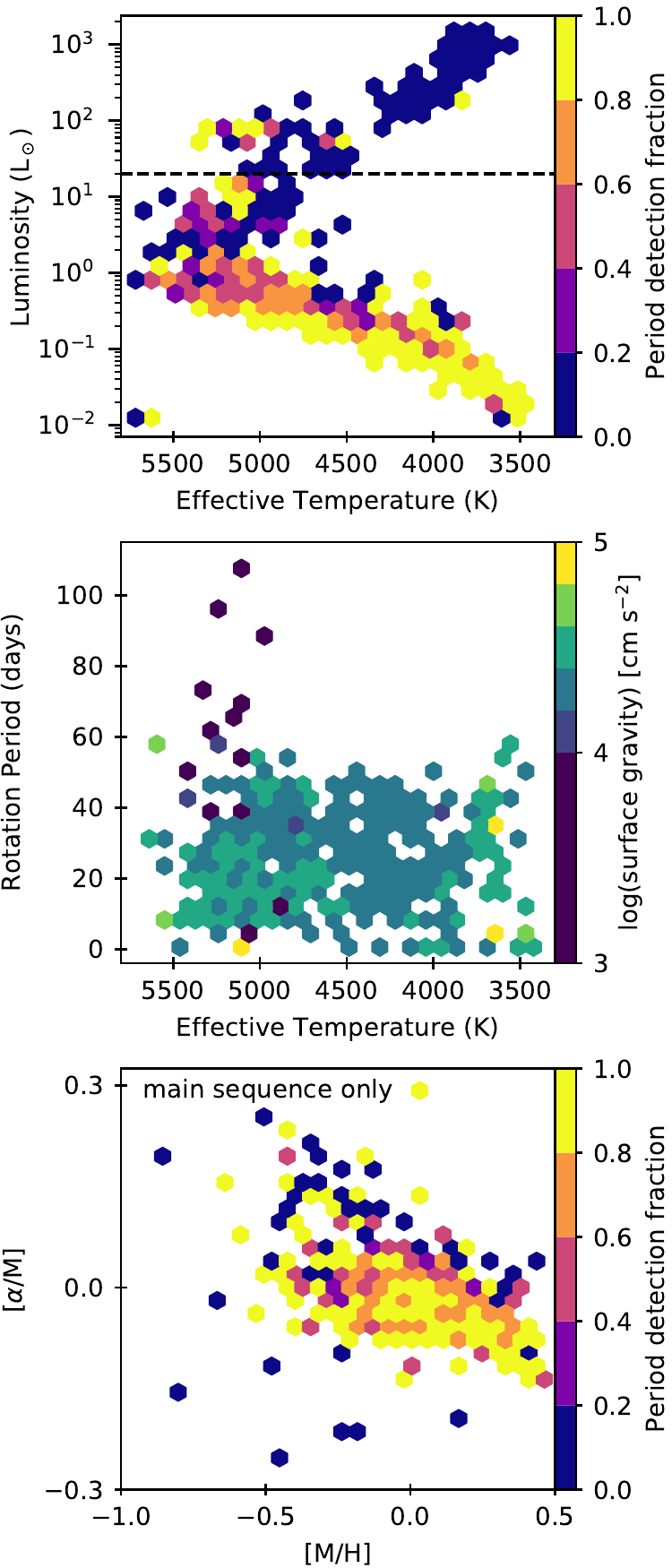}
    \caption{\textit{Top:} H-R diagram for our sample of cool \textit{Kepler} stars. We discarded giant-branch stars with $L > 20L_\odot$, shown by the black dashed line. \textit{Middle:} Rotation period vs. temperature for stars with reliable period detections. We detected rotation in only a handful of evolved stars. \textit{Bottom}: $\alpha$-abundance vs. metallicity showing the old, ``thick" disk ([$\alpha$/M] $\gtrsim$ 0.1) and young, ``thin" disk ([$\alpha$/M] $\lesssim$ 0.1).}
    \label{fig:star_diagnostics}
\end{figure}

The three panels of Fig.~\ref{fig:star_diagnostics} show where the stars in our sample are located on the H-R diagram, rotation--temperature space, and composition space. Following the method of \citet{Garcia14b}, \citet{Ceillier16, Ceillier17}, and \citet{Santos19}, we derived rotation periods from \textit{Kepler} light curves obtained with KADACS \citep[\textit{Kepler} Asteroseismic Data Analysis and Calibration Software,][]{Garcia11} using three high-pass filters at 20, 55, and 80 days and interpolating gaps in the data using the inpainting techniques of \citet{Garcia14a} and \citet{Pires15}. The rotational analysis employs wavelet decomposition based on \citet{Torrence98} and as implemented in the A2Z pipeline \citep{Mathur10}, autocorrelation function \citep[ACF, e.g.,][]{McQuillan14}, and the product of the two \citep[composite spectrum, CS; e.g.,][]{Ceillier17}, obtaining three period estimates for each target, one for each KADACS filter. \citet{Aigrain15} used injection/recovery methods to test the accuracy of this and other period-detection methods; they found that between 88\% (noisy simulated stars) to 92\% (noise free stars) of the periods recovered using this technique were accurate to within 10\%. Targets with reliable rotation-period estimates are automatically selected if the different estimates agree and the heights of the ACF and CS peaks are larger than a given threshold \citep[for details see][]{Santos19}. For the remainder of the sample, we proceed with visual inspection. The rotation estimate that we provide and use for the subsequent analysis is that retrieved from the wavelet decomposition. Of the \obsstars{} stars in our sample, we recovered rotation period estimates for \reliableperiods{} stars, $65\%$ being automatically selected. The remaining stars were discarded. We detected rotation periods for more stars using this method than using autocorrelation alone: based on the detection fractions in \citet{McQuillan14}, we would expect $\sim450$ detections. Our period estimates have a median uncertainty of $\sim2$ days and a median relative uncertainty of $\sim8\%$. 

Periods were not reliably detected in the remaining stars for a number of reasons, likely due to noise, unfavorable orientations of the rotation axes, long rotation periods (many tens to hundreds of days), and/or small amplitudes of spot modulation. The amplitude of the spot modulation depends on the stellar inclination angle and spot latitudes, but it also depends on spot area coverage which is expected to depend on stellar age. Thus, both period and amplitude of the spot modulation are markers of age, so we suspect that we preferentially detect rotation for young stars. This is evident in the bottom panel of Fig.~\ref{fig:star_diagnostics}, where most of the metal-poor, $\alpha$-rich stars (which tend to be old) are preferentially undetected in rotation.

We adopted temperatures, metallicities ([M/H]), and $\alpha$-element abundances ([$\alpha$/M]; $\alpha$ includes O, Mg, Si, S, Ca, and Ti, but is dominated by Mg, O, and Si in the dwarfs) from APOGEE Data Release 14 \citep{Holtzman18}. These parameters were derived from spectra by the APOGEE Stellar Parameters and Chemical Abundances Pipeline \citep[ASPCAP,][]{GarciaPerez16}. ASPCAP performs a temperature-dependent calibration to the abundance measurements under the assumption that stars in a given open cluster should have homogeneous abundances. We used these calibrated values for 461 of our stars with detected rotation periods; the remaining 71 stars had no calibrated spectroscopic parameters available, usually due to the proximity of the stars' temperatures or metallicities to ASPCAP atmosphere grid edges. For these stars we used the uncalibrated values. The temperatures have a median uncertainty of 100 K, while the metallicities and $\alpha$-abundances have median formal uncertainties of 0.06 and 0.02 dex, respectively. The spectroscopic temperatures are, on average, about 20 K cooler than their photometric counterparts reported by \citet{Berger18}, with a root-mean-square scatter of 147 K. These offsets are consistent with those reported by \citet{Holtzman18} for dwarfs in APOGEE DR14. \citet{Serenelli17} noticed a dispersion in ASPCAP metallicity when compared to values derived from optical spectroscopy; following their practice, we added a 0.1 dex uncertainty in quadrature to the ASPCAP formal uncertainties in both [M/H] and [$\alpha$/M]. For the uncalibrated parameters where uncertainties were not reported, we adopted 100 K uncertainty in temperature and 0.1 dex uncertainty in both metallicity and $\alpha$-abundance.

As an extra parameter to constrain the evolutionary states of stars in our sample, we adopted luminosities derived by \citet{Berger18} from \textit{Gaia} parallaxes. Our sample contains 31 stars which \citet{Berger18} are missing, and for these we used ASPCAP-derived surface gravities to constrain evolutionary phase.

Single red giants are predicted to have very long rotation periods, and detected rotation signals in them are typically related to binary mass transfer \citep{Tayar15}. We therefore focus on main-sequence period--age relations, discarding stars with luminosities above $20L_\odot$ from our analysis. We retain subgiants to gauge the usefulness of our braking law for non-main-sequence stars. Of the stars with no luminosity information, we removed those with $\logg < 3$. Using the luminosity and surface gravity cuts, \evocut{} stars were removed from our sample. Finally, 40 of the remaining stars were common to the giant candidate catalog of Garc\'ia et al.~(in prep). These stars' light curves and periodograms were visually inspected to ensure there was no contaminating power excess at the frequencies from which we derived rotation periods. Upon visual inspection, another \rafacut{} stars were cut, leaving \numMCMC{} for further analysis. 

Because our original sample was selected using KIC surface gravities, the giant contamination represents cases in which the original KIC was ineffective at separating evolved and unevolved stars. The fraction of contaminating giants and subgiants in our original sample of \obsstars{} stars is 25\%. This increases to 45\% when we consider stars redder than $K_p - J = 2$, in contrast to 74\% found in \citet{Mann12} for stars with KIC-determined $\logg > 4$, \textit{Kepler} magnitude $K_p < 14$, and $K_p - J > 2$. The difference is likely due to the magnitude cut in our sample (7 $< H$-magnitude $<$ 11): our faint-end limiting magnitude results in a smaller observed volume and less giant contamination than \citet{Mann12} observed.

\subsection{Markov Chain Monte Carlo (MCMC)}
To obtain ages, we used Markov Chain Monte Carlo sampling to map the stellar observables of metallicity, temperature, luminosity, and period onto the fundamental parameters of mass, bulk composition, and age. We ran chains for \numMCMC{} stars using the Python \texttt{emcee} package developed by \citet{Foreman-Mackey13}. \texttt{emcee} samples a posterior probability distribution using an ensemble of walkers, each running its own Markov chain. 

For priors, we used uniform constraints on mass, metallicity, and $\alpha$-abundance only to ensure the sampler remained within our model grid edges. For the log-likelihood, we used a $\chi^2$ in the form of:
\begin{align}
    &\ln{P_\mathrm{like}} = -\frac{1}{2}(\chi^2 + \chi_\mathrm{evo}^2) \nonumber \\
    &\chi^2 = \sum_i^N{\left(\frac{x_{\mathrm{obs}, i} - x_{\mathrm{mod}, i}}{\sigma_{x_{\mathrm{obs}, i}}}\right)^2}, \nonumber
\end{align}
where the $x_{\mathrm{obs}, i}$ are the \textit{Kepler} rotation period and ASPCAP-derived $\Teff$, [M/H] and [$\alpha$/M]; and the $x_{\mathrm{mod}, i}$ are the sampled YREC model parameters. The second term $\chi_\mathrm{evo}^2$ was used as a relatively weak constraint on the evolutionary state, and only when the sampled model luminosity was different from the observed luminosity by more than a factor of two. It took the form:
\begin{align}
    \chi_\mathrm{evo}^2 =
    \begin{dcases} \nonumber
        \left(\frac{\frac{1}{2} L_\mathrm{obs} - L_\mathrm{mod}}{\sigma_{L_\mathrm{obs}}}\right)^2, &L_\mathrm{mod} < \frac{1}{2} L_\mathrm{obs} \\
        ~~~~~~~~~~~0, &\frac{1}{2} L_\mathrm{obs} \leq L_\mathrm{mod} \leq 2L_\mathrm{obs} \\
        \left(\frac{2 L_\mathrm{obs} - L_\mathrm{mod}}{\sigma_{L_\mathrm{obs}}}\right)^2, &L_\mathrm{mod} > 2L_\mathrm{obs}
    \end{dcases}
\end{align}
This form encouraged the walkers to remain within a factor of two of the observed luminosity without allowing luminosity the same constraining power as rotation period and temperature. Thus, if the period and luminosity provided contradictory constraints (e.g., in the case of an unresolved wide binary system, the observed luminosity would be greater than that of a single star, but the observed period may be uncontaminated), the period information would dominate the likelihood value. 

An approximation was necessary to make the composition data compatible with the model grid we used. Namely, about half of our stars had $\alpha$-abundances that were less than solar (with the lowest being [$\alpha$/M]$= -0.44$), but the \citet{Castelli04} atmosphere tables are available only for [$\alpha$/M] $ = 0$ or +$0.4$. Rather than extrapolate to subsolar values, we modeled any star with an $\alpha$-abundance less than solar using a solar value. Since $\alpha$-enhancement has an effect on the angular momentum loss, this will have induced a slight bias in our determined ages for these stars. We discuss this bias in Sect.~\ref{sec:bias}.

\section{Results}
\label{sec:Results}
For the MCMC sampler, we used 12 walkers, a burn-in phase of 300 iterations, and a final sampling of 5000 iterations. We allowed our chains to be run to at least 50 autocorrelation times to ensure convergence to a single region of our sampling space. We also attempted runs using up to 50 walkers and 10,000 iterations for ten stars to test convergence and found our estimates with shorter chains to be robust.

\begin{deluxetable}{l|lll}
    \caption{Offsets of Markov chain median values to corresponding input values. Statistics were computed only for stars whose chains we labeled as ``good" (see Sect.~\ref{sec:Results}). For $\alpha$-abundance, the statistics were computed using only stars with positive input values, as stars with negative input values have biased chains due to the grid edge in [$\alpha$/M].}
    \label{tab:offsets}
    \tablehead{& \colhead{mean} & \colhead{median} & \colhead{RMedS}}
    \startdata
        $\Delta T_\mathrm{eff}$ (K)    & 23   & 16     & 16  \\
        $\Delta${[}M/H{]} (dex)        & 0.02 & 0.02   & 0.03  \\
        $\Delta${[}$\alpha$/M{]} (dex) & 0.05 & 0.06   & 0.06  \\
        $\Delta P_\mathrm{rot}$ (days) & 0.17 & 0.12   & 0.14 \\ 
    \enddata
\end{deluxetable}

We obtained chains for the \numMCMC{} stars in our sample. If the input values of temperature, rotation period, metallicity, and $\alpha$-abundance (for positive [$\alpha$/M] only) were within the 68\% credible interval of the chain, we labeled the chain as ``good". With these criteria, \numgood{} stars received the label of ``good". We computed the mean, median, and root-median-square offset (RMedS) between the input parameters and the median value of their corresponding chains for ``good" stars; these values are shown in Table~\ref{tab:offsets}. For all input parameters, the mean and median offsets are approximately equal to the root-median-square values, which we take to represent the median scatter about zero. In addition, the scatter in each of the four parameters is smaller than the median uncertainty in the input quantity. These demonstrate that the offsets between the input parameters and their respective model parameters are consistent with zero, indicating that our chains are well-converged to their input parameters.

\begin{figure}
    \centering
    \includegraphics[width=\linewidth]{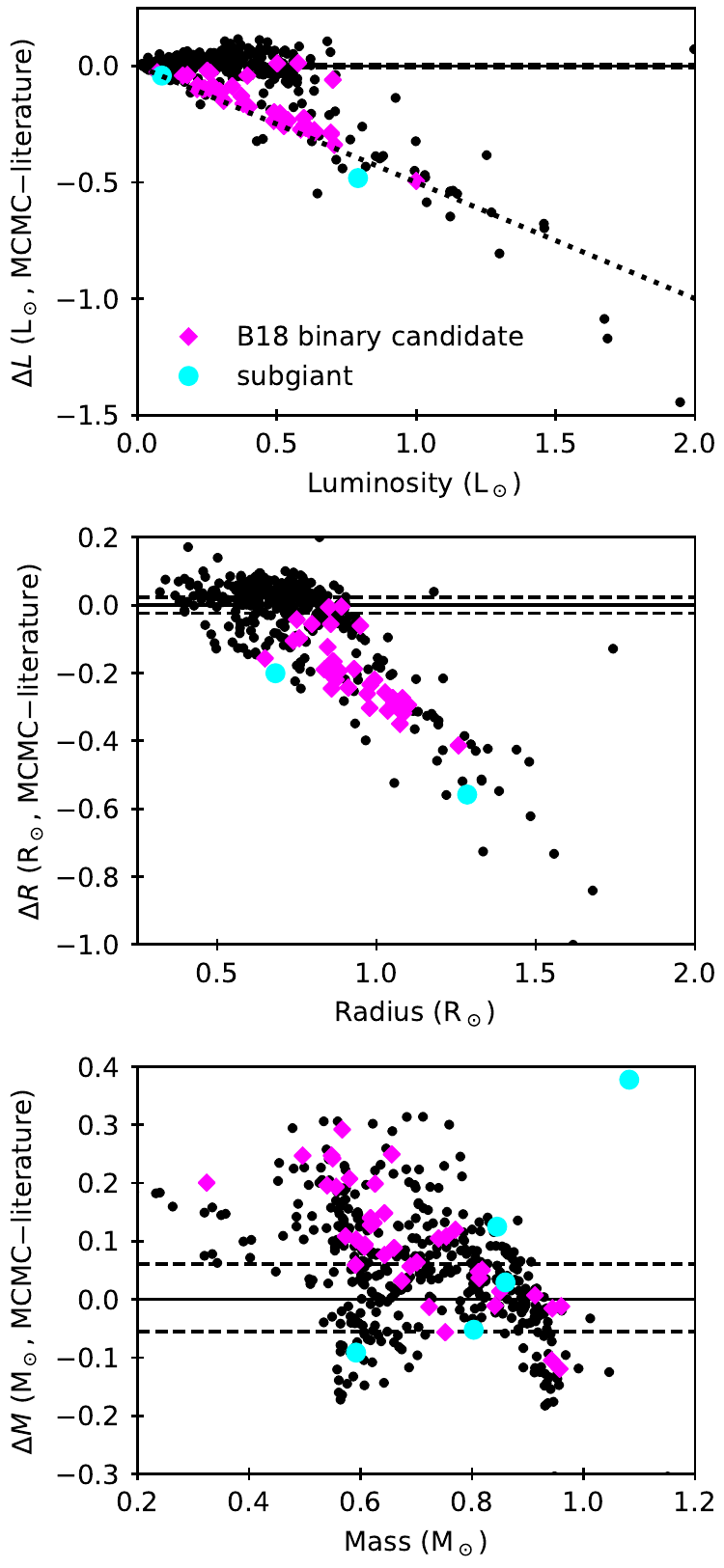}
    \caption{Comparison of inferred stellar parameters to literature values. Differences are given as posterior median minus the literature value; the dashed lines represent the median uncertainty in the literature values. The dotted line in the top panel represents $\Delta L/L = 50$\%, which is the maximum offset expected if the system is truly an equal-mass binary. The literature values of luminosity and mass are from \citet{Berger18} and \citet{Mathur17}, respectively, while the comparison radii were derived from luminosity and ASPCAP temperature. The 11 subgiants were reasonably well fit, but are out of view on the luminosity and radius panels.}
    \label{fig:convergence_plots}
\end{figure}

Given a value for each of these input quantities, a model provides other useful predictions as well. For example, we obtained as auxiliary quantities luminosity, radius, and mass, which provided another useful check for convergence and internal consistency. Fig.~\ref{fig:convergence_plots} depicts the Markov chain medians of these quantities compared to values from the literature. The most obvious feature is the group of stars that fall along the line $\Delta L = -\frac{L_\mathrm{lit}}{2}$, i.e., $L_\mathrm{lit} = 2L_\mathrm{model}$. This line is where we would expect equal mass binaries to fall, and \citet{Berger18} photometrically classified some ($\sim$ 30\%) of the stars along this line to be binary candidates, as marked by the magenta squares in the figure. The $\sim$ 30 black points along the same line may be unconfirmed binaries or stars with flux pollution from another source. Since the radii are derived from the luminosities, the same feature is seen in the radius plot.

Masking the stars confirmed as binaries, we computed statistics for the predicted luminosities, radii, and masses in a similar manner as for the input parameters. The remaining \numNotBinary{} stars have a median scatter about the one-to-one line of 0.01$L_\odot$, 0.01$R_\odot$, and 0.05$M_\odot$ in luminosity, radius, and mass, respectively. These values are comparable to the claimed uncertainties in the literature. For the stars with a Rossby number greater than that of the Sun (for which weakened braking may be important), the median scatter in $L$, $R$, and $M$ are $0.03L_\odot$, $0.03R_\odot$, and $0.06M_\odot$. Since these parameters were not directly used in the likelihood computation, the proximity of the chain medians to the literature values further reassures that our chains are converged.

\subsection{Gyrochronological Ages}

\input{data_table.tex}

\begin{figure*}
    \centering
    \includegraphics{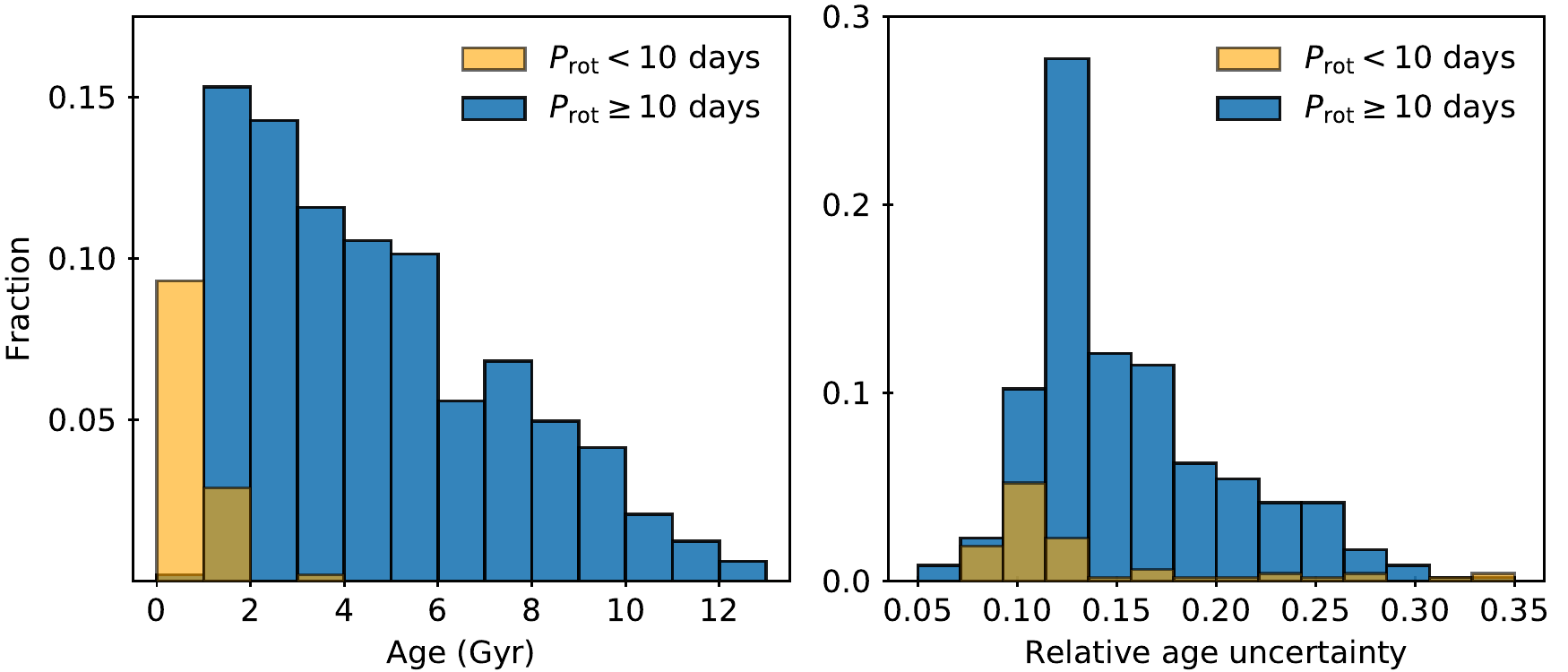}
    \caption{\textit{Left}: Distribution of median ages for ``good" stars. Stars with rotation periods less than ten days are likely to be synchronized binaries according to \citet{Simonian19}, so they are shown separately in orange. \textit{Right}: The distribution of relative age uncertainty. The median age uncertainty is 14\%, which is better than the median age uncertainties for other age determination methods on comparable samples.}
    \label{fig:age_histograms}
\end{figure*}

In sampling effective temperature, rotation period, metallicity, and $\alpha$-abundance, we obtained posterior distributions of ages for our sample of \numgood{} cool \textit{Kepler} dwarfs. We report for the first time gyrochronological ages inferred using full evolutionary models. The ages and other relevant stellar parameters for our sample are listed in Table~\ref{tab:resultsrot}. We consider the ages in this table to be reliable given our model assumptions. The distribution of ages in our reliable-period sample as well as the distribution of relative uncertainties are shown in Fig.~\ref{fig:age_histograms}. 

\citet{Simonian19} found that more than half of stars with observed periods between 1.5 and 7 days are actually synchronized binaries, noting that 10 days is the boundary longer than which no synchronization should take place \citep[but see also][]{Fleming19}. For this reason we plot stars with periods shorter than 10 days separately from those with longer periods.

Unlike the case of solar analogs, we clearly predict a large number of old ages for our targets, with no obvious break in the distribution. The age distribution peaks around 1 to 2 Gyr, which is younger than that found by \citet{Haywood13} and \citet{Buder19}, but consistent with the result of \citet{SilvaAguirre18}, who also studied stars in the \textit{Kepler} field. Even if the fast rotators are not binned separately, the peak remains unchanged. The median relative age error is 14\%, which is better than the 20--30\% range usually seen for isochrone ages of similar stellar populations \citep[e.g.,][who studied over 7000 dwarf and main-sequence-turnoff stars in the Galactic Archaeology with HERMES survey, GALAH]{Buder19}. For a discussion of systematic bias in our age estimates, see Sect.~\ref{sec:bias}.

We do not expect our sample to be representative of the distribution underlying the galactic stellar population for several reasons. First, our selection criteria remove stars that have KIC surface gravity $\logg < 4$ dex. Since these stars are evolved and therefore preferentially older, this cut biases our sample toward younger stars. Moreover, requiring that our stars' rotation be detectable biases the sample toward more active stars, which also tend to be younger. Furthermore, although the original sample was selected with simple cuts in magnitude, temperature, and surface gravity in the KIC, the KIC itself is somewhat biased in its determination of stellar parameters.

\subsection{A word on rotational detection bias}

\begin{figure}
    \centering
    \includegraphics[width=\linewidth]{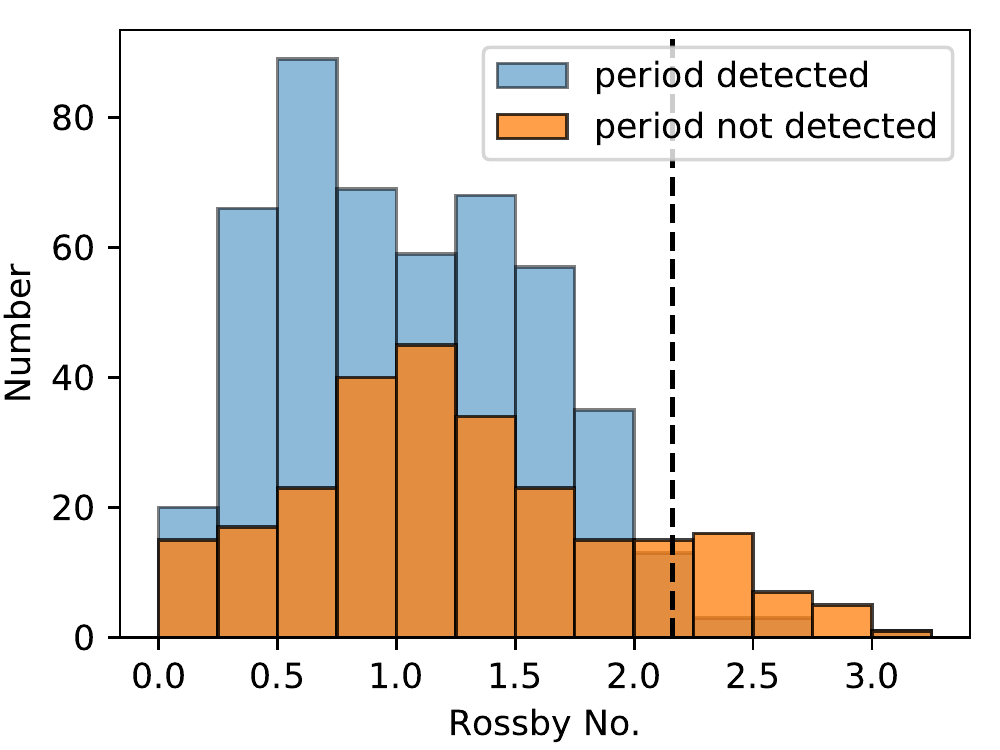}
    \caption{Observed Rossby numbers for the rotationally detected (blue) and estimated Rossby numbers for the undetected (orange) stars on the main sequence. The vertical dashed line is $\textrm{Ro} = \textrm{Ro}_\odot = 2.16$. Below this line, we expect to detect all stars in rotation with exceptions for unfavorable inclinations and spot configurations. Above it, we expect stars' variability to fall below detection thresholds.}
    \label{fig:rossby_histogram}
\end{figure}

Our sample of stellar ages is uniquely biased in comparison to other such galactic samples: our ability to infer ages is tied directly to our ability to detect rotational modulation. This modulation is stronger in younger, more rapidly rotating stars. At fixed rotation period, it is stronger in stars with cooler temperatures and deeper convective envelopes \citep[see][]{McQuillan14, vanSaders19}. $\alpha$-enhancement also tends to deepen convective envelopes at fixed metallicity, although it is a second-order effect in comparison to the bulk metallicity \citep{Karoff18}. Taken together, it means that we expect our sample to be biased against old, metal poor objects, which is clearly seen in the bottom panel of Fig.~\ref{fig:star_diagnostics}. This is in contrast to asteroseismology, which is not particularly biased against old stars, and with isochrone methods, which tend to struggle with young and intermediate age stars on the main sequence in this temperature range.

\citet{vanSaders19} suggested that stars with Rossby number $\textrm{Ro}>2$ might become undetectable in spot modulation based on the solar-temperature stars in the \citet{McQuillan14} field sample, creating a sharp ``detection edge" at long periods. For the metallicities and $\alpha$-enhancements of stars undetected in spot modulation in our sample, we would have expected all stars cooler than 4500 K to fall at $\textrm{Ro}< 2$ for $t<12$ Gyr, and thus to fall below this detection edge. 

 For those stars with measured rotation periods, we calculated the Rossby number $(\textrm{Ro}= P_\mathrm{rot}/ \tau_\mathrm{cz})$ using the observed period and the model convective overturn timescale from our MCMC stellar parameter estimation. To estimate a distribution of Rossby numbers for the stars \textit{not} detected in rotation we used the following recipe: 
\begin{enumerate}
    \item{We split the stars into an $\alpha$-rich and $\alpha$-poor sequence using the same [M/H]--[$\alpha$/M] selection as \citet[see their Fig.~8]{SilvaAguirre18}.}
    \item{We randomly sampled an age for each star from the appropriate $\alpha$-rich or $\alpha$-poor age distribution of \citet[see their Fig.~9]{SilvaAguirre18}.}
    \item{Using the star's ASPCAP abundances, sampled age, and mass from \citet{Mathur17}, we interpolated a Rossby number for each star from our stellar model grid.}
\end{enumerate}
Fig.~\ref{fig:rossby_histogram} shows a histogram of Rossby numbers for the rotationally undetected stars, compared with the MCMC-estimated Rossby numbers for our ``good'' sample. While we can say little about the robustness of a Rossby number estimate for an individual undetected star, statistically the set of Rossby numbers should provide some insight into the sample of rotationally undetected stars. Strikingly, the undetected stars are not overwhelmingly predicted to have high Rossby numbers: the undetected stars, like the detected stars, mostly have $\textrm{Ro} < 2$, save a small but notable tail of undetected stars with predicted Rossby numbers $\textrm{Ro}>2$. This suggests that the non-detections are only mildly biased against the oldest stars; most are likely undetected due to unfavorable spot patterns or inclinations. 

\begin{figure}
    \centering
    \includegraphics[width=\linewidth]{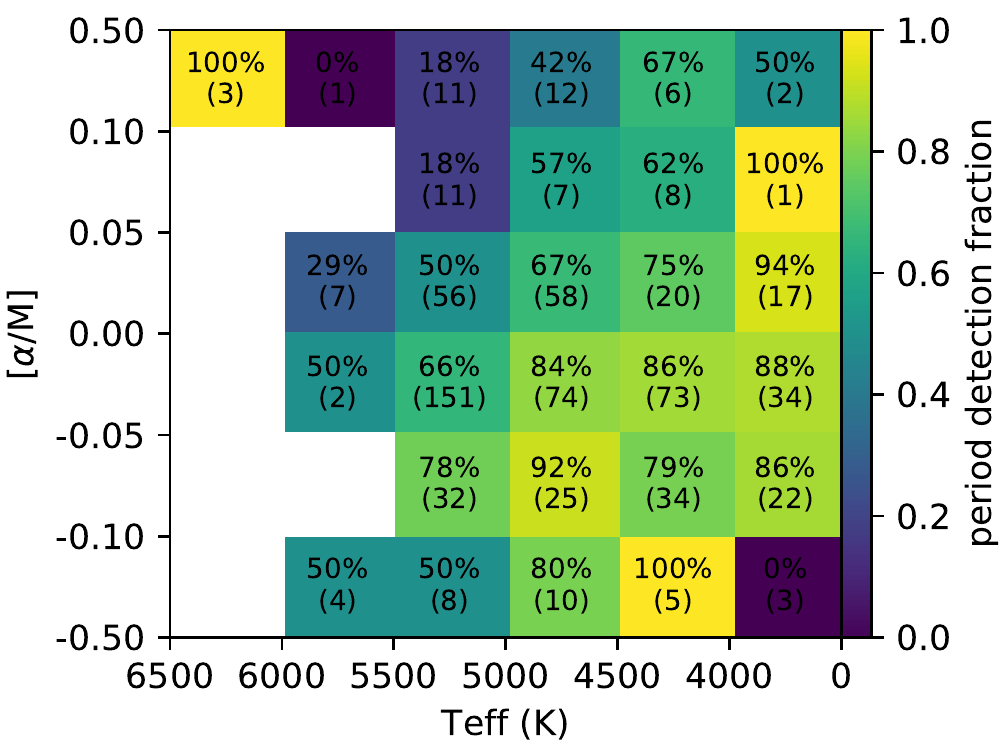}
    \caption{Rotation detection fraction as functions of temperature and $\alpha$-enhancement for stars with $\logg > 4$. Color denotes the fraction of stars detected in rotation in each bin, while each bin is labeled with both the detection fraction and the number of stars contained in that bin. Like \citet{McQuillan14}, we find better detectability at cooler temperatures, but we also find $\alpha$-enhanced stars are more difficult to detect in rotation, presumably due to the link between $\alpha$-enhancement with age.}
    \label{fig:detection_grid}
\end{figure}

Another view of detectability is encapsulated in Fig.~\ref{fig:detection_grid}, which shows a 2D histogram of the fraction of stars detected in rotation as a function of effective temperature and $\alpha$-enhancement. We include all stars in our sample with ASPCAP $\logg > 4$ so that the fractions are unaffected by contamination by giant stars. Like \citet{McQuillan14}, we find higher detection fractions at cooler temperatures. The decreasing detectability with increasing $\alpha$-enhancement represents the difficulty of detecting rotation in progressively older and more slowly rotating stars.

\subsection{Chemical Evolution Trends}

Here we examine whether we recover the various chemical enrichment trends observed in studies that use ages determined from isochrones or asteroseismology. 

\subsubsection{Trends in \texorpdfstring{$\alpha$}{}-abundance with age}
\label{sec:alpha-age-trends}
\begin{figure*}
    \centering
    \includegraphics[width=\linewidth]{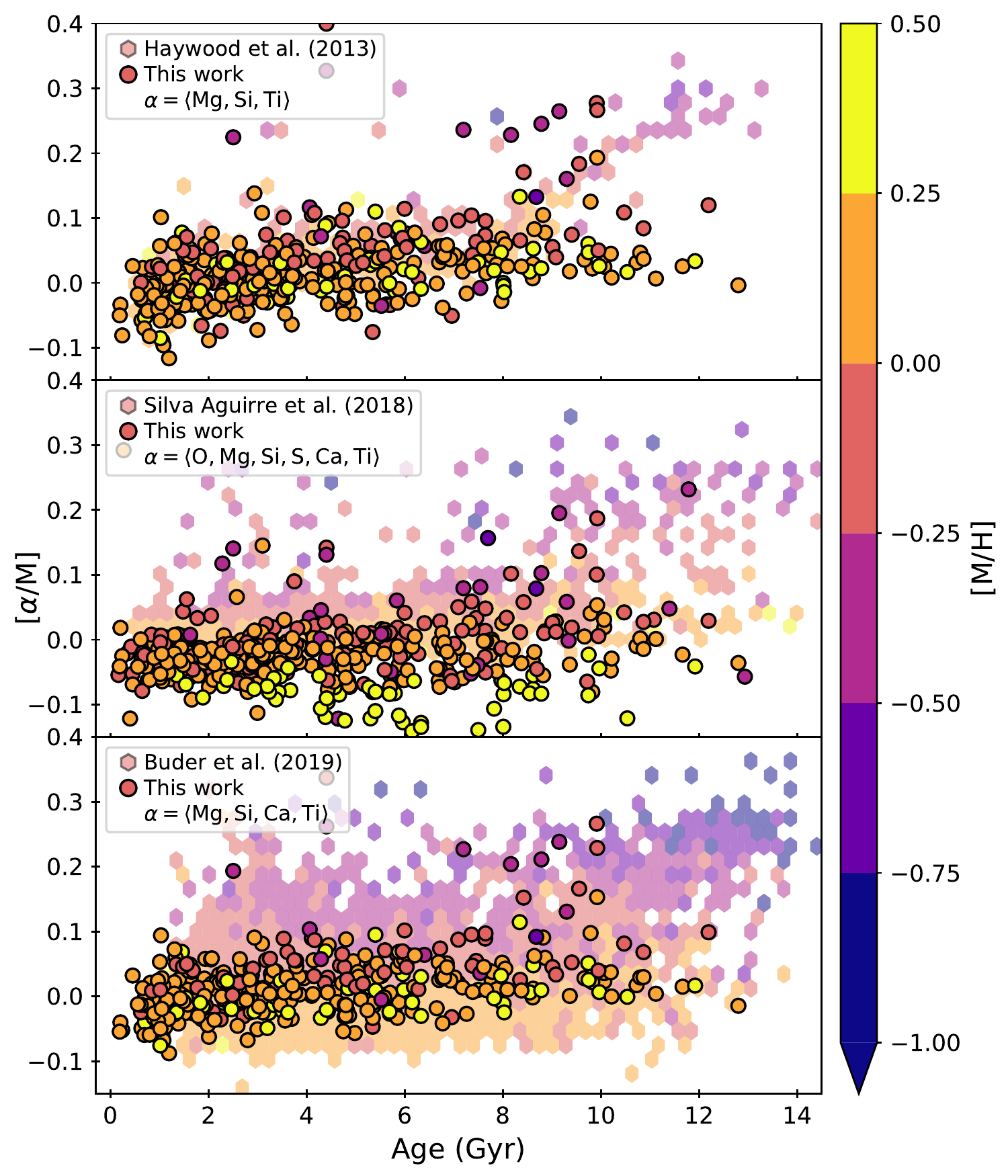}
    \caption{The [$\alpha$/M] vs. age distribution for our ``good" stars (shown in circular points), compared with results from other work \citep[][shown in hexagonal bins]{Haywood13, SilvaAguirre18, Buder19}. We recover the $\alpha$-enhancement slope for stars younger than 8 Gyr seen by \citet{Haywood13}, but we also find old, $\alpha$-poor stars seen by \citet{Buder19} and \citet{SilvaAguirre18} that \citet{Haywood13} did not see. 
    Our lack of old, $\alpha$-enhanced stars is likely due to observational biases (see Sect.~\ref{sec:alpha-age-trends}).}
    \label{fig:alpha_age_results}
\end{figure*}

Using our rotation-based ages, we investigated the abundance of $\alpha$-elements as a function of age to look for trends in galactic chemical evolution. Our results are shown in Fig.~\ref{fig:alpha_age_results} and are plotted over results of the similar studies of \citet{Haywood13}, \citet{SilvaAguirre18}, and \citet{Buder19} to compare age trends. Each comparison study uses a different combination of $\alpha$-elements; to ensure accurate comparisons, we plot our ages against the appropriate combination of elemental abundances from APOGEE. We consider trends of individual $\alpha$-elements with age in Appendix~\ref{sec:appendix}.

We recover the general trend of increasing $\alpha$-abundance with age. The rate of $\alpha$-enhancement is consistent with the rates observed by \citet{Haywood13} and \citet{SilvaAguirre18} for stars younger than $\sim 8$ Gyr, but the sharp upturn in slope at about 8 Gyr seen by \citet{Haywood13} is absent from our sample. Instead, we observe a spread of $\alpha$-abundances for old stars more consistent with the results of \citet{SilvaAguirre18} and \citet{Buder19}. The difference is likely due to the sample selection used by the different studies: \citet{Haywood13} selected stars in the solar vicinity, while \citet{SilvaAguirre18} and \citet{Buder19} study stars in APOGEE and GALAH, which probe beyond the solar neighborhood and detect a wider variety of stars.

\begin{figure}
    \centering
    \includegraphics[width=\linewidth]{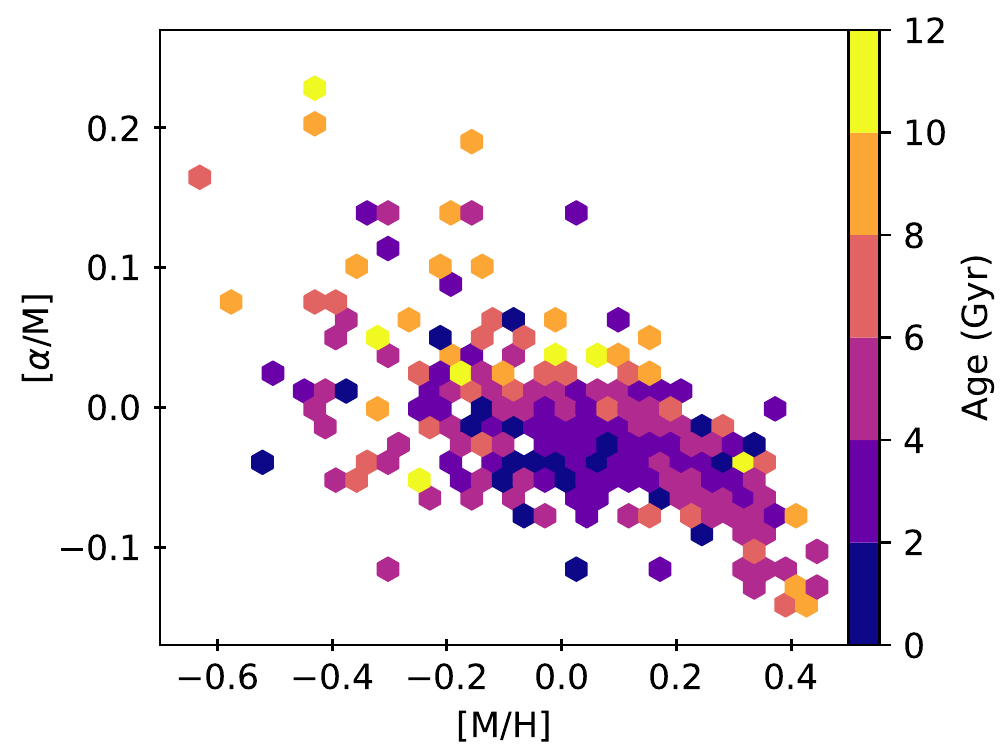}
    \caption{The same composition-space diagram as the bottom of Fig.~\ref{fig:star_diagnostics}, except here the bins are colored by rotation-based age. As expected, most of the $\alpha$-rich ([$\alpha/M$] $\gtrsim 0.1$) stars are fairly old. We emphasize that the lack of stars in the $\alpha$-enhanced disk is mainly due to two observational biases: (1) the stars \textit{Kepler} observed in the $\alpha$-rich, thick-disk sequence were usually evolved and showed little or no rotational modulation in their light curves, and (2) older main-sequence stars tend to be less magnetically active and are consequently more difficult to detect in rotation.}
    \label{fig:met_alpha_age}
\end{figure}

Perhaps the most drastic difference between our sample and the comparison samples is that we find very few stars in the old, $\alpha$-enriched sequence. Fig.~\ref{fig:met_alpha_age} shows this same result in composition space, this time colored by age. The lower, young, $\alpha$-poor sequence is well populated, but the upper $\alpha$-rich sequence is nearly empty. This is likely due to the fact that old, $\alpha$-enhanced stars are relatively inactive and are therefore difficult to detect in rotation (see again the bottom of Fig.~\ref{fig:star_diagnostics}). Metallicity may play a role too: old stars tend to be relatively metal poor and thus have thinner convective envelopes. Presumably, this would result in weaker magnetic fields, less starspot activity, and fewer detections at fixed age. In addition, almost all the stars in our sample are located within 0.5 kpc of Earth. The $\alpha$-rich ``thick" disk stars tend to be farther away, mostly at distances greater than 1 kpc \citep[see Fig.~4 of][]{Hayden15}, well outside of \textit{Kepler}'s peak detectability. 

\subsubsection{Young, \texorpdfstring{$\alpha$}{}-rich stars}
Our sample includes \numyoungalpharich\ young ($< 6$ Gyr), $\alpha$-rich ($\textrm{[}\alpha\textrm{/M]} > 0.13$) stars, selected to have periods $P > 10$ days. These stars are present in numerous other surveys \citep{Martig15,Chiappini15,SilvaAguirre18,Buder19} where ages are derived by a variety of other means including isochrone fitting, asteroseismology, and the [C/N]--mass relationship in giants. \citet{Chiappini15} suggested that this population is formed in the region near corotation with the Galactic bar, and is then subject to strong radial migration. They may also represent a blue straggler population \citep{Martig15}. We caution that if these stars are indeed merger products, our rotation-based ages are unreliable: if mass transfer spins up the stars, they will appear young when ages are inferred through rotation--age relations. Interestingly, these stars are also unusually kinematically hot (see Fig.~\ref{fig:kinematics}).

\subsubsection{Old, \texorpdfstring{$\alpha$}{}-poor stars}
\citet{Buder19} found that $30\%$ of the old (age $> 11$ Gyr) stars were $\alpha$-poor ([$\alpha$/Fe] $<$ 0.125). Our sample is even more strikingly $\alpha$-poor: \fracalphapoor\% of stars older than 11 Gyr have low $\alpha$ abundances. We suspect that this is largely due to the sample bias discussed above. The old stars in this sample are predominantly $\alpha$-poor, but they are also metal-rich ($-0.25 <$ [M/H] $< 0.45$). We suspect that the rotational detection bias preferentially removes old, metal-poor stars from the sample, resulting in an apparent enrichment of old, metal-rich, $\alpha$-poor stars. 

\subsubsection{Metallicity dispersion}
We find that the dispersion in metallicity increases for older stars, similar to \citet{Buder19} and \citet{SilvaAguirre18}. We bin our ``good" sample with periods greater than 10 days in 1.5 Gyr bins (each with more than 20 stars), and find that the scatter in metallicity increases from \metscatteryoung\ at 2.0 Gyr to \metscatterint\ at 5.0 Gyr, to \metscatterold\ at 9.5 Gyr, consistent with the increased scatter found in stars with isochrone ages in \citet{Buder19}.  

\subsubsection{Metallicity distribution}
The distribution of metallicities in our sample corroborates the suspicion of detection bias: while our full sample (including stars for which we did not detect rotation) has $\langle$[Fe/H]$\rangle = -0.001 \pm 0.008$, our ``good" sample is more metal-rich with $\langle$[Fe/H]$\rangle = 0.055 \pm 0.009$ (a 4.5-$\sigma$ difference). The mean and difference are the same for [M/H] as for [Fe/H]. For comparison, the full solar neighborhood sample of APOGEE Data Release 12 \citep{Hayden15} has $\langle$[Fe/H]$\rangle = 0.01 \pm 0.002$, while the sample of \citet{Buder19} is more metal-poor with $\langle$[Fe/H]$\rangle$ = $-0.04 \pm 0.003$. The slight difference in mean metallicity between the APOGEE and GALAH samples is likely due to their respective target selections. APOGEE targeted the (metal-rich) galactic disk, while GALAH avoided it \citep{Buder19}.

\subsection{Kinematic trends}

We examine the kinematics of our sample as a function of age using the combination of positions, proper motions, parallaxes, and radial velocities from the second \textit{Gaia} data release \citep{Gaia18}. We draw associate \textit{Kepler} targets with their \textit{Gaia} counterparts following the method of \citet{Berger18}, which compared the magnitudes of the \textit{Kepler} and \textit{Gaia} source matches, removed stars with uncertain ($\sigma_{\pi}/\pi > 0.2$) parallaxes, and implemented the astrometric quality cuts described in detail in \citet{Berger18}, \citet{Lindegren18}, and \citet{Arenou18}. In total, \gaiaRVsample\ stars in our sample survive the \citet{Berger18} cuts, have reported \textit{Gaia} RV values \citep{Gaia18}, and have a ``good" label. We convert the \textit{Gaia} five-parameter astrometic solution and line of sight velocities into U (radial), V (azimuthal), and W (vertical) velocities following \citet{Johnson87}. We adopt a peculiar velocity for the Sun of $(U_{\odot}, V_{\odot},W_{\odot}) = (11.0,12.24,7.25)$ km/s from \citet{Schoenrich10}. We estimate velocity errors by propagating the full \textit{Gaia} 5D $(\alpha, \delta, \varpi, \mu_\mathrm{ra}, \mu_\mathrm{dec})$ covariance matrix through the coordinate transformations, assuming an independent Gaussian error distribution for the line of sight velocities.

\begin{figure}
    \centering
    \includegraphics[width=\linewidth]{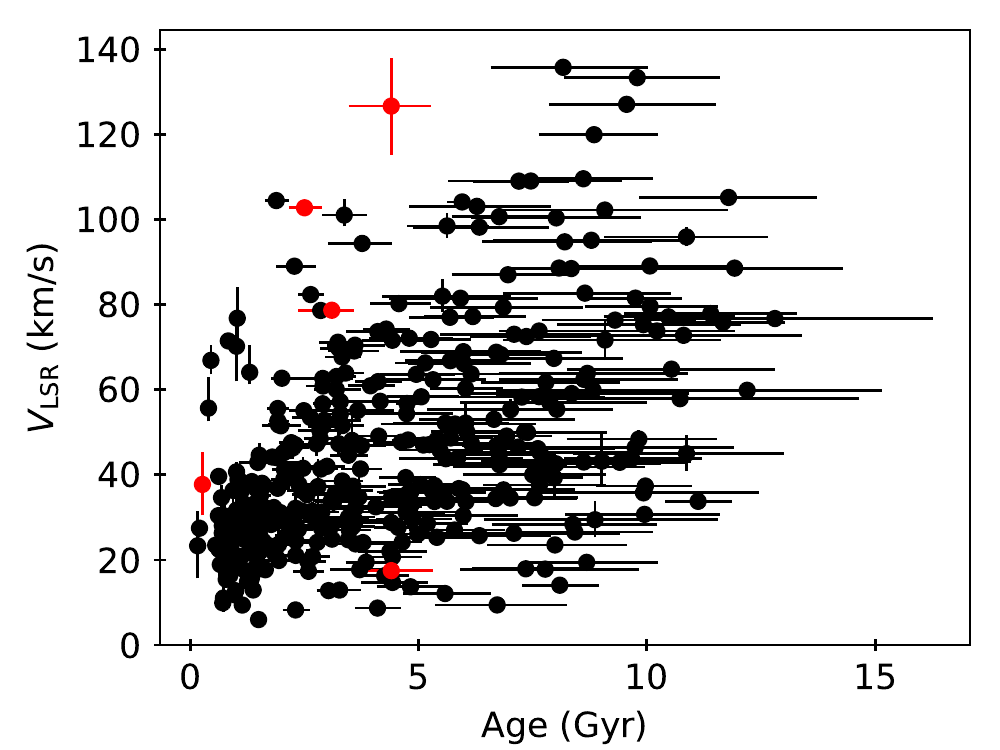}
    \caption{Velocities with respect to the Local Standard of Rest ($V_\mathrm{LSR} = \sqrt{U_\mathrm{LSR}^2 + V_\mathrm{LSR}^2 + W_\mathrm{LSR}^2}$) for sample stars with ``good" designation, displaying increased velocity dispersion with age. Young $\alpha$-enhanced stars ($t<6$ Gyr, $\textrm{[}\alpha/\textrm{M]} > 0.13$) are shown in red. All stars are shown, regardless of rotation period.}
    \label{fig:kinematics}
\end{figure}

We show the velocities with respect to the local standard of rest in Fig.~\ref{fig:kinematics} as a function of rotation-based age. The velocity dispersion in the sample is higher in the older stars, consistent with a picture in which stars are dynamically heated over time. Our sample shows much the same behavior as the M-dwarf sample of \citet[][their Fig.~13]{Newton16}, who also observed an increase velocity dispersion in slow rotators, corresponding to old stars. 

\section{Discussion}
We have inferred rotation-based ages for this cool-star sample in part because all other standard age determination methods fail in dwarf stars much cooler than the Sun. Even in our models, magnetic braking is calibrated using only a handful of stars with $\Teff < 5500$K at ages greater than a few Gyr. Our inferred ages are therefore still an extrapolation of relations calibrated on largely solar-like objects. The recovery of broad kinematic and galactochemical trends seen in samples with ages determined via other means is an encouraging sign that rotation does indeed carry age information in cool stars. 

While the ages we present have a median relative uncertainty of 14\%, we note that this relates to precision, not accuracy. This precision comes under the assumption that our choice of braking law is correct, and under a particular set of input physics for the stellar models. With these assumptions, the \textit{relative} ages of our stars are likely to be accurate, but we can make no such claim for the absolute ages. Fortunately, investigations of evolution generally rely on relative ages, so accurate absolute ages are not always necessary.

A systematic look at the effects of different stellar model grids, magnetic braking models, and angular momentum transport prescriptions is beyond the scope of this work. However, here we do examine the impacts of a range of different initial rotation periods and the manner in which composition is incorporated into the modeling. 

\subsection{Systematic uncertainties in gyrochronological ages}
\label{sec:bias}

The uncertainties associated with model-based gyrochronological ages include effects of composition and initial conditions. For composition, changing the bulk metallicity or detailed abundances changes the depth of the stellar convection zone, which in turn affects the rate of spin-down. Meanwhile, it has been demonstrated that a star ``forgets" its initial conditions after some sufficient amount of time has passed, usually less than a Gyr for G- and K-dwarfs, but up to several Gyrs for the coolest M-dwarfs \citep[e.g.,][]{Epstein14, Gallet15}. Any lingering dependence on initial conditions remains to be quantified. Using our model grid, we performed three tests to quantify independently the effects of metallicity, $\alpha$-abundance, and initial conditions on our derived ages.

First we investigated the effects on age if metallicity is ignored, i.e., if one assumes a star has solar composition. We attempted to mimic the scenario one might encounter when attempting to use gyrochronology relations on a poorly studied field star whose temperature and period are known, but whose metallicity, detailed abundances, or initial rotation period are not. We employed the following procedure:
\begin{enumerate}
\item{Interpolate ages for two stars from the model grid using the same randomly sampled mass, $\alpha$-abundance, and rotation period, but with different metallicities: one with [M/H] = 0.5, and one with a solar value (i.e., [M/H] = 0)}
\item{Compare the ages for the two stars, treating the metal-rich star as the ``real" star, and the star with solar abundance as an ``incorrect'' model.}
\item{Repeat until all regions of the model grid have been sufficiently sampled}
\end{enumerate}
We repeated this test for a metal-poor case as well, using a ``real" star with [M/H] = $-0.5$.

The second test was like the first except that the $\alpha$-enhancement was used as the manipulated variable. In other words, we sampled two stars of the same mass, metallicity, and rotation period, but differing $\alpha$-abundances of [$\alpha$/M] = 0 (the solar case) and 0.4 (the enhanced case).

Finally, in our third test we sampled stars of the same mass, metallicity, and $\alpha$-abundance, but with different initial conditions. The control sample employed the standard ``fast-launch" conditions of disk-locking timescale $\tau_\mathrm{disk} = 0.281$ Myr and disk period $P_\mathrm{disk} = 8.134$ days. The test sample, on the other hand, used ``slow-launch" conditions of $\tau_\mathrm{disk} = 5.425$ Myr and $P_\mathrm{disk} = 13.809$ days \citep{vanSaders13}. These values were calibrated using stars in the Pleiades cluster (125 Myr old) and M37 (550 Myr). 

\begin{figure*}
    \centering
    \includegraphics{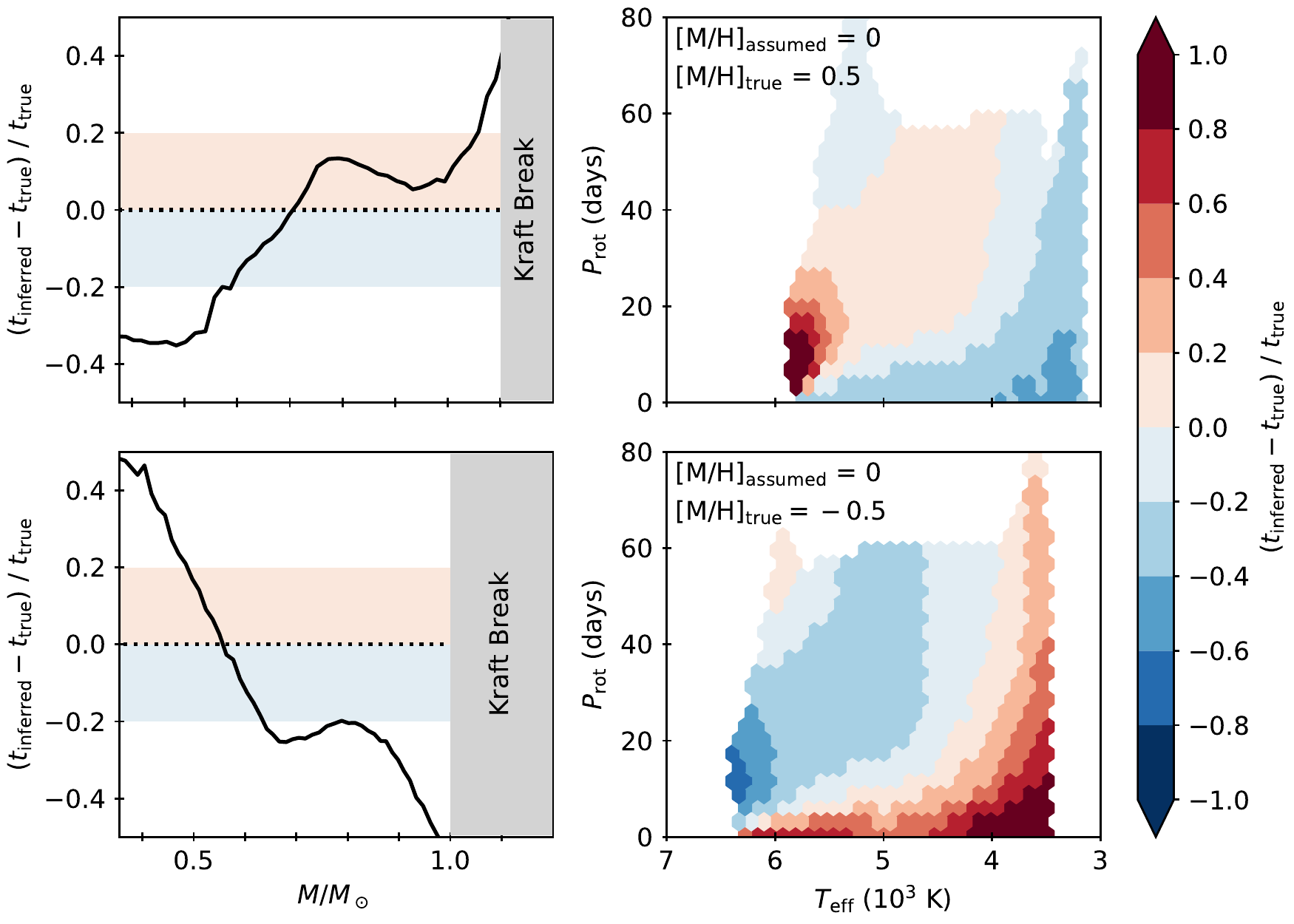}
    \caption{Fractional bias in inferred age when solar metallicity is incorrectly assumed and the actual metallicity is 0.5 dex (top) and $-0.5$ dex (bottom). \textit{Left:} Age bias changes as a function of mass. Below a solar mass, the effect can range from $-40$\%--$10\%$ in metal-rich stars, but the effect is slightly worse for metal-poor stars and at the Kraft break. \textit{Right:} The same age bias now shown on the $P_\mathrm{rot}$--$\Teff$ diagram. The bias is worst in M-dwarfs (where calibration is poor) and stars near the Kraft break (where the use of gyrochronology is ill-posed).}  
    \label{fig:met_tests}
\end{figure*}

\begin{figure*}
    \centering
    \includegraphics{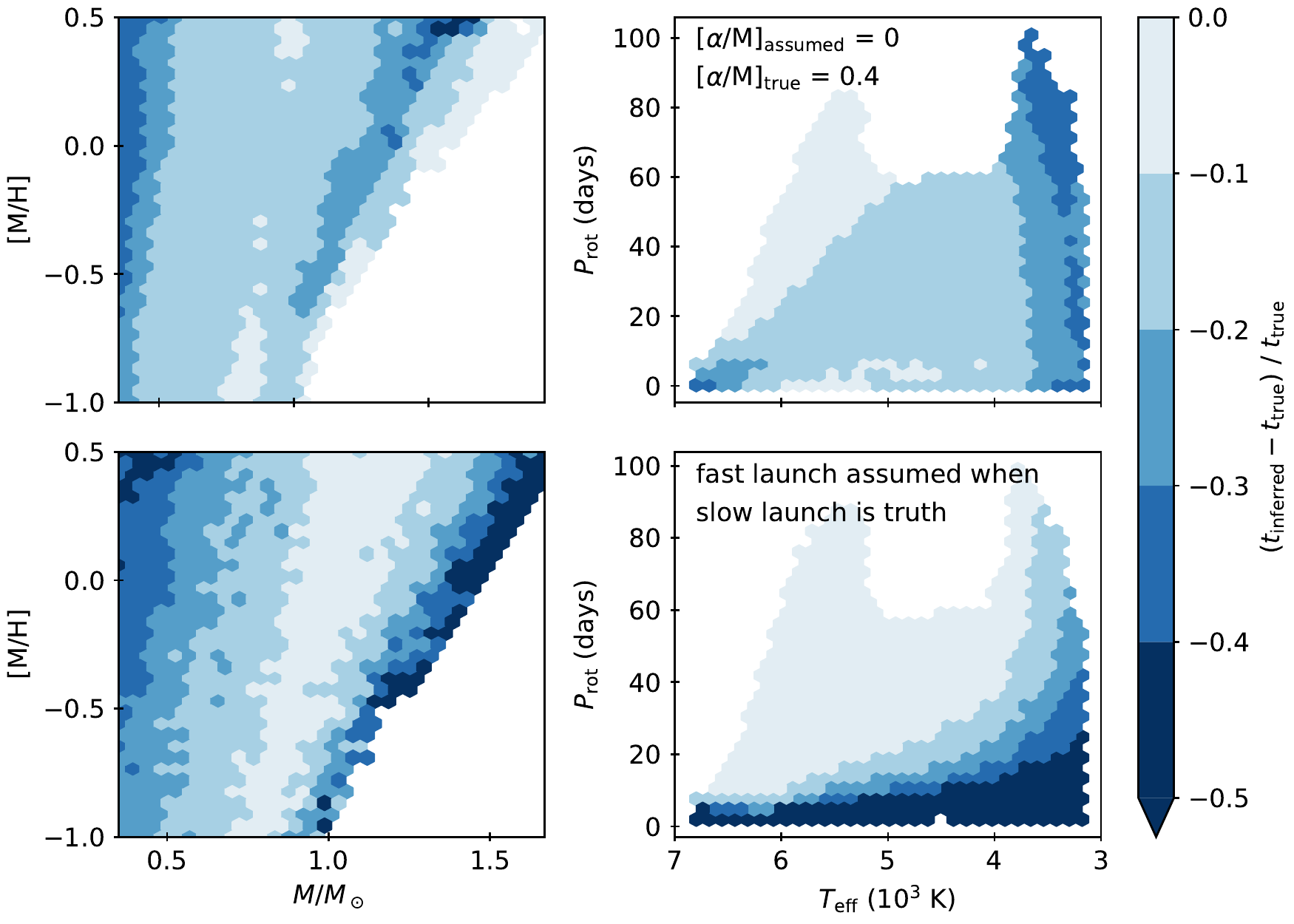}
    \caption{\textit{Top}: Fractional bias in inferred age when $\alpha$-abundance is incorrectly assumed to be solar and the actual enhancement is 0.4 dex. For most of the mass range where gyrochronology is validated, ignoring $\alpha$-enhancement can lead to 10--20\% biases. \textit{Bottom}: Fractional bias in inferred age when fast-launch initial conditions are assumed, but the actual initial conditions are slow-launch. For late G- and K-dwarfs, the bias is less than 10\%, but the error is worse for M-dwarfs, which take longer to forget their initial rotation periods.}
    \label{fig:alpha_launch_tests}
\end{figure*}

Each test consisted of 50,000 random samples, and the results are shown in Figs.~\ref{fig:met_tests} and \ref{fig:alpha_launch_tests}. In all plots, the value $(t_\mathrm{inferred} - t_\mathrm{true})/t_\mathrm{true}$ indicates by what fraction the age is overestimated (or underestimated) under the given false assumption. The panels in a given row show the results for the same sample in two different ways. The left panels of Fig.~\ref{fig:met_tests} show the age bias as a function of mass, while the right panels show a $\Teff$--period diagram with bins colored by the age bias. For the $\alpha$-abundance and launch condition tests, the bias depended on metallicity, so the left panels of Fig.~\ref{fig:alpha_launch_tests} have metallicity on the vertical axis and are colored by age bias.

If one assumes solar metallicity for a metal-rich, $0.75\mathrm{M}_\odot$ star (top of Fig.~\ref{fig:met_tests}), the age will be overestimated by $\sim$ 15\%. The bias is much larger for stars more massive than the Sun, i.e., near the Kraft break. Here a subtle change in metallicity changes whether the star has a convective envelope and spins down over time, so the inferred rotation-based age is drastically affected. These effects are slightly worse, but in the opposite direction, for the metal-poor case: for a metal-poor, 0.75$M_\odot$ star, the age will be underestimated by $\sim 40\%$. In addition, the Kraft break occurs at lower masses in metal-poor stars, so the bias is also worse than the metal-rich case for stars slightly less massive than the Sun. On the low-mass end, the bias is reversed: for a metal-rich M-dwarf, assuming solar metallicity results in \textit{underestimating} the age by $\sim$ 30\%. However, this is also the regime in which a spread of initial rotation periods persists. Both of these cases are beyond regimes where gyrochronology has been validated: above the Kraft break, the lack of spin-down makes any kind of rotation-dating useless, while M-dwarfs suffer from a lack of calibrators. If we consider the somewhat better-studied regime of 0.5--1.0$M_\odot$, the bias ranges from $-30\%$ (for metal-poor stars) to 15\% (for metal-rich stars) when comparing the inferred ages of two stars of identical mass and period, ignoring metallicity. For the observer's case where temperature and period are known (say, for example, they are solar) and metallicity is assumed to be solar, the inferred age for a truly metal-rich ([M/H] = 0.5) star will be overestimated by $\sim$ 15\%, while the inferred age for a truly metal-poor ([M/H] = $-0.5$) will be underestimated by $\sim$ 30\%. Similarly, two stars with identical $\Teff = 4500$K and $P_\mathrm{rot} = 40$ days will have ages different by $\sim$ 5\% in the metal-rich case and by $\sim$ 10\% in the metal-poor case. Metallicity is therefore an important consideration in full evolutionary spin-down models; more calibrators at a range of metallicities should be pursued.

The top row of Fig.~\ref{fig:alpha_launch_tests} shows the results of the $\alpha$-abundance test. If one assumes solar $\alpha$-abundance for an $\alpha$-enhanced, solar-metallicity star of 0.75$M_\odot$, detailed abundances enter the systematic error budget at the 10--20\% level in a model grid. This bias remains fairly constant in the mass ranges where gyrochronology is useful, but the bias is more drastic in stars near the Kraft break, similarly to the metallicity tests. This suggests that even if the metallicity is known perfectly, rotation-based ages may be biased by as much as 20\% if detailed abundances are not known to some level. Since the atmosphere tables used in our model grids are defined only for [$\alpha$/M] = 0 and 0.4, we cannot say without extrapolation how this bias behaves in $\alpha$-poor stars. 

Finally, the bottom row of Fig.~\ref{fig:alpha_launch_tests} shows the bias in age if fast-launch models are used, but the system is truly slow-launch. For late G- and K-dwarfs, assuming the incorrect launch condition generally does not result in more than a 10\% bias in age because sufficient time has passed to allow them to ``forget" their initial rotation conditions. For late K- and early M-dwarfs, incorrectly assuming initial conditions comes into the systematic error budget at the $\sim$ 10--20\% level, with the bias growing worse toward M-dwarfs, which take longer to forget their initial conditions. The bias is also larger for stars near the Kraft break; since stars in this regime do not spin down with age, two identical stars with differing launch conditions never have the same period on the main sequence. This results in an undefined age bias, as shown by the blank bins in the bottom-left panel of Fig.~\ref{fig:alpha_launch_tests}.

\subsection{Other underlying assumptions}
While we have tested the sensitivity of our assumed braking law to initial conditions and metallicity, we have neglected a number of confounding factors. We assume only one form of the period--age relation. While all period--age relations reproduce the behavior of solar-temperature stars by construction, they tend to make diverging predictions at cooler temperatures \citep[see Fig.~1 in][]{Matt15}. This is largely a consequence of the scarcity of empirical calibrators for cool stars. The asteroseismic calibrator sample extends only to $\sim0.8 \textrm{ M}_{\odot}$ \citep{vanSaders16} and open cluster sequences are also truncated at roughly the same mass in systems older than $\sim1$ Gyr \citep{Meibom15,Agueros18,Curtis19}.  A detailed assessment of systematic age offsets produced by different literature braking laws is beyond the scope of this article. Here instead we have focused on whether under the assumption of a \textit{particular} braking law we can recover the galactochemical trends inferred using other age metrics. 

Furthermore, we have neglected the impact of internal angular momentum transport in our models. To maintain solid-body rotation, the transport in the interior must be fast in comparison to the angular momentum loss from the surface, which is not necessarily true during all phases of evolution. In so called ``two-zone" models \citep{MacGregor91}, the core and envelope are treated as separate zones, with angular momentum transport occurring between them on a characteristic timescale $\tau_\textrm{c-e}$. Two-zone models can display slower rotation than the solid body case (at fixed age) while the core has had insufficient time to couple (the effective angular momentum reservoir being depleted by magnetized winds is smaller), but faster than solid body rotation once the core and envelope recouple. Recoupling timescales in Sun-like stars are $\tau_\textrm{c-e} \approx$ 10--55 Myr \citep{Denissenkov10, Gallet15, Lanzafame15, Somers16}, while for 0.5$M_{\odot}$ stars $\tau_\textrm{c-e} \approx 10^2$--$10^3$ Myrs \citep{Gallet15, Lanzafame15}. During the preparation of this manuscript, \citet{Curtis19} published rotation periods for K- and M-dwarfs in the 1 Gyr old cluster NGC 6811, and showed that spin-down appears to stall between the ages of Praesepe ($\sim 700$ Myr) and NGC 6811 (1 Gyr), which they attribute to potential core-envelope decoupling. However, \citet{vanSaders19} showed that the longest-period stars from \citet{McQuillan14} with $\Teff < 5000$K have periods fully consistent with a braking law of the solid body form that we have adopted here for stars roughly the age of the galactic disk, suggesting that the impact of this stalled braking does not persist for old ages. The impact of core-envelope decoupling on the observed rotation periods in K- and M-stars is uncertain, due again to a lack of calibrators with known ages. While a detailed treatment of core-envelope decoupling is also beyond the scope of this article, we caution that if it is important, the age scale we present here is likely stretched or compressed with respect to the true underlying age scale, particularly at ages of hundreds of Myr (Sun-like stars) to several Gyr (M-dwarfs). 

\section{Summary and Conclusion}
\label{sec:Conclusion}
We have presented gyrochronological ages for stars from the APOGEE-\textit{Kepler} Cool Dwarf sample. In doing so, we have quantified sources of bias that must be considered when inferring ages from rotation, and we have reproduced chemical--age trends in the literature using an independent means of age determination. Our results are as follows:
\begin{enumerate}
    \item We provide rotation-based age estimates for \numgood{} stars (mostly cool dwarfs) in the \textit{Kepler} field. In contrast with gyrochronology results from solar analogs, we find a substantial population of stars older than 4 Gyr with no evidence for a sharp detection threshold below at least 10 Gyr.
    \item Our age distribution peaks at younger ages than seen in broader chronology studies, but it is consistent with studies of \textit{Kepler} giants.
    \item We recover the trend of increasing $\alpha$-abundance with age for young ($<$ 8 Gyr) stars seen in other studies. 
    \item We find a population of old ($>$ 8 Gyr), $\alpha$-poor stars seen in recent studies.
    \item We see a clear trend of increasing velocity dispersion in old stars. 
    \item We observe \numyoungalpharich{} young, $\alpha$-rich stars in our sample. Because mass transfer would affect rotation periods, we cannot differentiate between whether these are truly young, single, $\alpha$-enhanced stars or merger/interaction products. However, these stars are kinematically hot, making them more consistent with an older population, which may favor the merger hypothesis.
    \item There is tentative evidence for a drop in detection efficiency for $\alpha$-rich stars relative to $\alpha$-poor ones, which may indicate that there is an age bias in gyrochronology for stars older than 10 Gyr, even on the lower main sequence.
    \item We find that ignoring metallicity effects in gyrochronology can result in biases in age estimates as low as $-30\%$ for metal-poor stars and as high as $15\%$ for metal-rich stars.
    \item Similarly, ignoring detailed abundance patterns such as $\alpha$-enhancement can result in biases at the 10--20\% level.
    \item Initial conditions enter the systematic error budget below the 10\% level for late G- and early K-dwarfs, but at $\sim 10$--20\% for late K- and early M-dwarfs.
\end{enumerate}

Our results demonstrate the power of gyrochronology as a tool for problems like galactic evolution. They also underscore the assumptions inherent in rotation-based age inference and the observational biases associated with detecting rotation based on stellar spot modulation. The limits of our study highlight the need for larger and deeper samples if rotation continues to be used to probe the history of the Milky Way. While \textit{Kepler} revolutionized the study of stellar rotation, its pointing near the galactic disk meant that the farthest (and oldest) reaches of the galaxy remain poorly probed. The \textit{Transiting Exoplanet Survey Satellite} \citep[\textit{TESS},][]{Ricker15}, due to the orientation of its continuous viewing zones out of the galactic disk, will probe further into the $\alpha$-enhanced disk than \textit{Kepler}. Consequently, \textit{TESS} will enable gyrochronological study for a much broader population of stars.

Together, the abundance of precise rotation periods from \textit{Kepler} and large spectroscopic datasets such as APOGEE provide valuable resources for inferring stellar ages. Whereas not long ago, isochrone-based techniques offered the most reliable age estimates for large sets of stars, we have now entered an era in which rotation-based ages can compete with and surpass the precision and feasibility of other chronometers. This will be essential in understanding evolution on a galactic scale, which requires detailed knowledge of age with chemistry, kinematics, and position for immense numbers of stars.

\acknowledgements
This work involved the development of \texttt{kiauhoku}, a stellar model grid interpolator built for interacting with YREC models. This name was created in partnership with Dr. Larry Kimura and Bruce Torres Fischer, a student participant in \textit{A Hua He Inoa}, a program to bring Hawaiian naming practices to new astronomical discoveries. We are grateful for their collaboration.

We thank Jennifer Johnson for her effort in constructing this cool dwarf sample, and for comments that improved the quality of this manuscript. 

We are grateful to Sven Buder for providing stellar ages from the preprint of his 2019 paper. We also thank Victor Silva-Aguirre for agreeing to supply the ages from his 2018 work.

This research was supported in part by the National Science Foundation under Grant No. NSF PHY-1748958.

JVS and ZRC acknowledge support from the TESS Guest Investigator Program (80NSSC18K18584).

ARGS acknowledges the support from the National Aeronautics and Space Administration (NASA) under Grant NNX17AF27G.

RAG acknowledges the funding received from the PLATO/CNES grant.

SM acknowledges support by the Spanish Ministry under the Ramon y Cajal fellowship number RYC-2015-17697.

JT acknowledges support provided by NASA through Hubble Fellowship grant \#51424 awarded by the Space Telescope Science Institute, which is operated by the Association of Universities for Research in Astronomy, Inc., for NASA, under contract NAS 5-26555

This work includes data collected by the \textit{Kepler} mission. Funding for the \textit{Kepler} mission is provided by the NASA Science Mission directorate.

Funding for SDSS-III has been provided by the Alfred P. Sloan Foundation, the Participating Institutions, the National Science Foundation, and the U.S. Department of Energy Office of Science. The SDSS-III web site is \url{http://www.sdss3.org/}.

SDSS-III is managed by the Astrophysical Research Consortium for the Participating Institutions of the SDSS-III Collaboration including the University of Arizona, the Brazilian Participation Group, Brookhaven National Laboratory, Carnegie Mellon University, University of Florida, the French Participation Group, the German Participation Group, Harvard University, the Instituto de Astrofisica de Canarias, the Michigan State/Notre Dame/JINA Participation Group, Johns Hopkins University, Lawrence Berkeley National Laboratory, Max Planck Institute for Astrophysics, Max Planck Institute for Extraterrestrial Physics, New Mexico State University, New York University, Ohio State University, Pennsylvania State University, University of Portsmouth, Princeton University, the Spanish Participation Group, University of Tokyo, University of Utah, Vanderbilt University, University of Virginia, University of Washington, and Yale University.

\software{KADACS \citep{Garcia11}, YREC \citep{Demarque08}, \texttt{emcee-3} \citep{Foreman-Mackey13}, \texttt{kiauhoku} (\url{www.github.com/zclaytor/kiauhoku})}

\bibliography{references}
\bibliographystyle{aasjournal}

\appendix

\section{Elemental Trends with Age}
\label{sec:appendix}
\begin{figure*}
    \centering
    \includegraphics{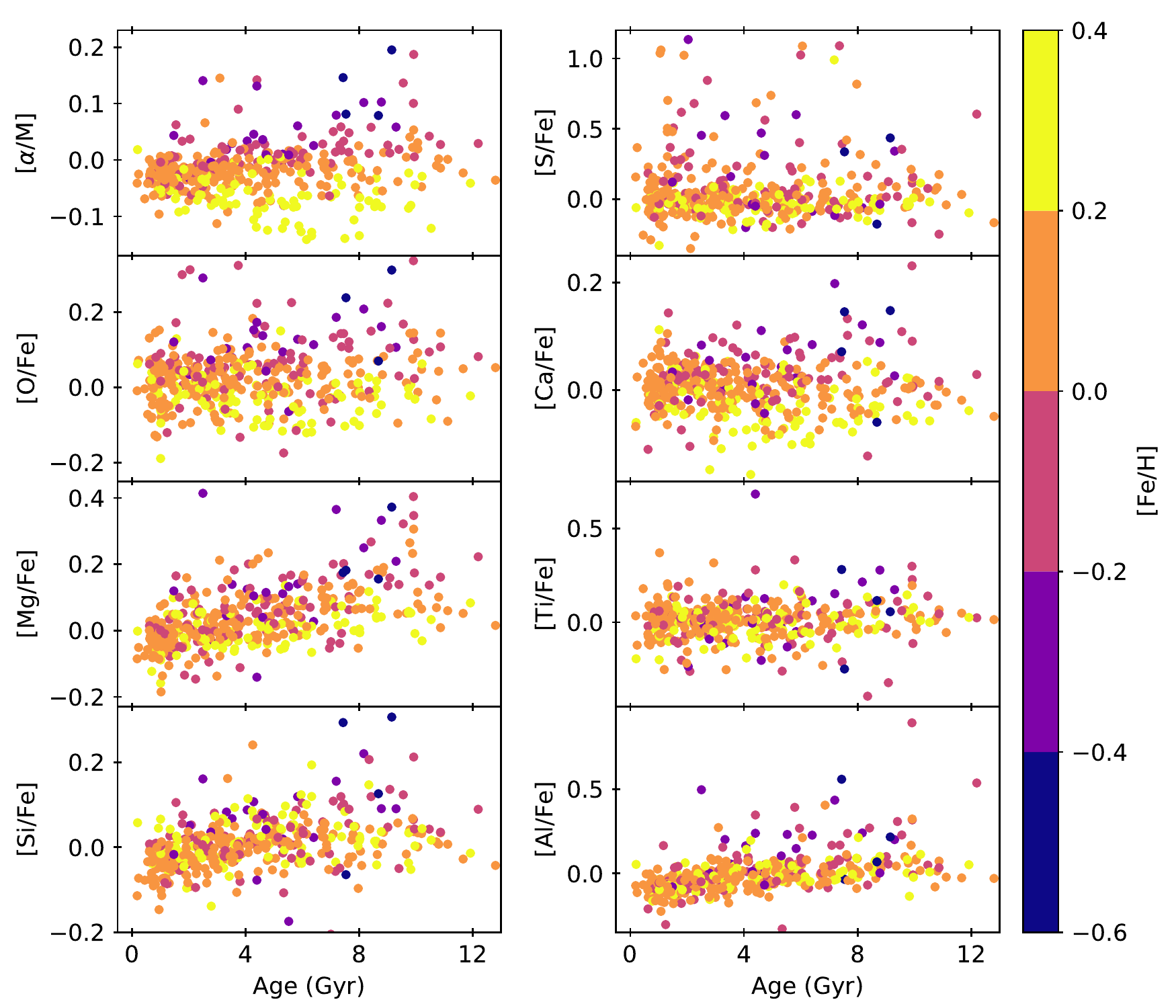}
    \caption{The enrichment of $\alpha$-process elements (and aluminum) with age.}
    \label{fig:elements}
\end{figure*}

APOGEE defines [$\alpha$/M] as the combination of O, Mg, Si, S, Ca, and Ti \citep{GarciaPerez16}. While [$\alpha$/M] should capture the general enrichment and depletion trends from different kinds of supernovae, we may consider the evolution of individual elements as well. Fig.~\ref{fig:elements} shows the evolution of the individual $\alpha$-elements, as well as aluminum, across time. While aluminum is not an $\alpha$-process element, we include it because its enrichment is generally driven by the same processes as $\alpha$-process elements \citep[e.g.,][]{Nomoto13}.

Magnesium and silicon show the strongest trends with age, which is expected since they, along with oxygen, dominate the value of [$\alpha$/M] in our dwarf sample. Interestingly, oxygen shows little-to-no trend with age. It does have the most scatter at fixed age; this scatter may blur out any trends across time.  

\end{document}

%% file: numbers.tex
\newcommand{\allstars}{1241}

\newcommand{\obsstars}{930}

\newcommand{\reliableperiods}{532}

\newcommand{\evocut}{15}

\newcommand{\rafacut}{15}

\newcommand{\numMCMC}{502}

\newcommand{\numgood}{483}

\newcommand{\numNotBinary}{415}

%% file: numbers_jvs.tex
\newcommand{\metscatteryoung}{\ensuremath{0.14}}
\newcommand{\metscatterint}{\ensuremath{0.22}}
\newcommand{\metscatterold}{\ensuremath{0.26}}
\newcommand{\fracalphapoor}{\ensuremath{89}}
\newcommand{\numyoungalpharich}{\ensuremath{3}}
\newcommand{\gaiaRVsample}{\ensuremath{381}}

%% file: data_table.tex


\begin{deluxetable*}{lcccccccccccccc}

\tablecaption{The input data and derived rotation-based ages for stars in the APOGEE-\textit{Kepler} cool dwarfs sample. We list the \textit{Kepler} IDs, APOGEE effective temperatures, bulk metallicities, and $\alpha$-enhancements with formal uncertainties, rotation periods with uncertainties estimated from \textit{Kepler} light curves, luminosities with uncertainties inferred from \textit{Gaia} parallaxes by \citet{Berger18},  and the median of the age posteriors with 68\% credibility limits.
\label{tab:resultsrot}}


\tablehead{\colhead{KIC ID} & \colhead{$\Teff$} & \colhead{$\sigma_{T}$} & \colhead{[M/H]} & \colhead{$\sigma_\mathrm{[M/H]}$} & \colhead{[$\alpha$/M]} & \colhead{$\sigma_\mathrm{[\alpha/M]}$} & \colhead{$P_\mathrm{rot}$} & \colhead{$\sigma_{P}$} & \colhead{$L$} & \colhead{$\sigma^{-}_{L}$} & \colhead{$\sigma^{+}_{L}$} & \colhead{Age} & \colhead{$\sigma_{\mathrm{age}}^{-}$} & \colhead{$\sigma_{\mathrm{age}}^{+}$} \\ 
\colhead{} & \colhead{(K)} & \colhead{(K)} & \colhead{(dex)} & \colhead{(dex)} & \colhead{(dex)} & \colhead{(dex)} & \colhead{(days)} & \colhead{(days)} & \colhead{($L_\odot$)} & \colhead{($L_\odot$)} & \colhead{($L_\odot$)} & \colhead{(Gyr)} & \colhead{(Gyr)} & \colhead{(Gyr)}} 

\startdata
1432745 & 4600 & 98 & 0.171 & 0.118 & -0.024 & 0.102 & 21.810 & 1.701 & 0.168 & 0.003 & 0.003 & 2.882 & 0.347 & 0.377 \\ 
1724975 & 5259 & 107 & -0.011 & 0.117 & -0.042 & 0.101 & 10.607 & 0.748 & 0.854 & 0.018 & 0.019 & 1.030 & 0.105 & 0.114 \\ 
1996721 & 5019 & 100 & -0.005 & 0.115 & 0.012 & 0.101 & 31.932 & 3.419 & 0.295 & 0.007 & 0.007 & 6.037 & 0.967 & 1.115 \\ 
2018047 & 4049 & 102 & -0.375 & 0.116 & 0.103 & 0.103 & 43.017 & 2.450 & 0.075 & 0.001 & 0.001 & 8.786 & 0.888 & 1.020 \\ 
2156061 & 4284 & 96 & 0.146 & 0.117 & -0.047 & 0.103 & 29.383 & 3.065 & 0.450 & 0.116 & 0.192 & 4.448 & 0.720 & 0.815
\enddata


\tablecomments{Table~\ref{tab:resultsrot} is published in its entirety in the machine-readable format. A portion is shown here for guidance regarding its form and content.}


\end{deluxetable*}